# On the critical competition between singlet exciton decay and free charge generation in non-fullerene based organic solar cells with low energetic offsets

Manasi Pranav[1], Atul Shukla[1], David Moser[2], Julia Rumeney[2], Wenlan Liu[3], Rong Wang[4], Bowen Sun[1], Sander Smeets[5,6], Nurlan Tokmoldin[1,7], Frank Jaiser[1], Thomas Hultzsch[1], Safa Shoaee[1,7], Wouter Maes[5,6], Larry Lüer[4], Christoph Brabec[4,8], Koen Vandewal[5,6], Denis Andrienko[3], Sabine Ludwigs[2], Dieter Neher[1*]

<u>Affiliations:</u>

[1] Institute of Physics and Astronomy, University of Potsdam, Karl-Liebknecht Straße 24/25, 14476 Potsdam, Germany

[2] IPOC – Functional Polymers, Institute of Polymer Chemistry, University of Stuttgart, Pfaffenwaldring 55, 70569 Stuttgart, Germany

[3] Max Planck Institute for Polymer Research, Ackermannweg 10, 55128 Mainz, Germany

[4] Institute of Materials for Electronics and Energy Technology (i-MEET), Friedrich-Alexander-Universität Erlangen-Nürnberg, Martensstrasse 7, Erlangen 91058, Germany

[5] UHasselt—Hasselt University, Institute for Materials Research, (IMO-IMOMEC), Agoralaan 1, 3590 Diepenbeek, Belgium

[6] IMOMEC Division, IMEC, Wetenschapspark 1, 3590 Diepenbeek, Belgium

[7] Heterostructure Semiconductor Physics, Paul-Drude-Institut für Festkörperelektronik, Leibniz-Institut im Forschungsverbund Berlin e.V, Hausvogteiplatz 5-7, 10117 Berlin, Germany

[8] Helmholtz-Institut Erlangen-Nürnberg for Renewable Energies (HIERN), Forschungszentrum Jülich, Immerwahrstraße 2, 91058 Erlangen, Germany

*Corresponding author: neher@uni-potsdam.de

## Abstract:

In this era of non-fullerene acceptor (NFA) based organic solar cells, reducing voltage losses while maintaining high photocurrents is the holy grail of current research. Recent focus lies in understanding the manifold fundamental mechanisms in organic blends with minimal energy offsets – particularly the relationship between ionization energy offset ($\Delta IE$) and free charge generation. We quantitatively probe this relationship in multiple NFA-based blends by mixing Y5 and Y6 NFAs with PM6 of varying molecular weights, covering a 15% to 1% power conversion efficiency (PCE) range and a progression of $\Delta IE$. Spectroelectrochemistry reveals a critical $\Delta IE$ of approximately 0.3 eV, below which the PCE sharply declines. Transient absorption spectroscopy consistently reveals that a smaller $\Delta IE$ slows the dissociation of the NFA's local singlet exciton (LE) into free charges, albeit restorable by an electric field. Bias-dependent time delayed collection experiments quantify the free charge generation efficiency, while photoluminescence quantum efficiency measurements assess photocurrent loss from LE decay. Combined with transient photoluminescence experiments, we find that the decay of singlet excitons is the primary competition to free charge generation in low-offset NFA-based organic solar cells, with neither noticeable losses from charge-transfer (CT) decay nor evidence for LE-CT hybridization. Our experimental data align with Marcus theory calculations, supported by density functional theory simulations, for zero-field free charge generation and exciton decay efficiencies. We find that efficient photocurrent generation generally requires that the CT state is located below the LE, but that this restriction is lifted in systems with a small reorganization energy for charge transfer.

## Introduction

Organic solar cells (OSCs) have witnessed remarkable improvement in performance since the advent of low-bandgap non-fullerene acceptors (NFAs). When blended with an appropriate electron donor (D) to form a bulk heterojunction (BHJ), state-of-the-art NFA-based OSCs nowadays match their inorganic and perovskite competitors in terms of high short-circuit current densities ($J_{SC}$) and fill factors ($FF$), but lag behind with respect to their open circuit voltage ($V_{OC}$).[1]–[4] This is in part due to the need of an energy offset between the blend constituents at the D:A heterojunction, which must be large enough to dissociate the initially photogenerated local singlet exciton (LE) into an interfacial charge transfer (CT) state and eventually into free charges (see **Figure 1a**). Since in most state-of-the-art OSCs the difference of the ionisation energies ($\Delta$IE) is smaller than that of the electron affinities ($\Delta$EA), the $\Delta$IE critically determines voltage losses. Furthermore, $\Delta$IE of the neat components is often used as a first approximation of the LE-CT energetic offset; however, this simplification neglects other contributions to the offset such as the difference of the LE and CT binding energies,[5] the electrostatics at the D:A interfaces,[6] or differences in the molecular packing and conformation in the blend and especially at the D:A interface.[7], [8]

There is an ongoing debate about the minimum $\Delta$IE required to guarantee efficient free charge generation but also about the main reasons for the decreasing performance of low offset D:A blends.[9]–[12]. For example, Nakano *et al.* demonstrated a strict correlation between the free charge generation efficiency and the LE-CT energy offset. It was concluded that the performance of low-offset devices is limited by inefficient LE dissociation rather than CT separation.[13] Classen *et al.* arrived at a similar conclusion, observing efficient exciton dissociation for an $\Delta$IE as small as 50 meV.[14] Therein, an equilibrium model was postulated where a low energetic driving force for charge generation (the LE to CT transition) can be partially compensated for by a long lived LE state. In contrast, Karuthedath *et al.* and later Gorenflot *et al.* showed that an $\Delta$IE of at least 0.45 eV is needed to guarantee efficient hole transfer but also that the charge generation efficiency decreases rather gradual for decreasing $\Delta$IE.[9], [12] In addition to previous work, these authors also highlighted the role of band bending due to the large quadrupole moment of many NFAs, which lifts the CT state above the local LE in low-offset blends. In another work, Qian *et al.* probed a series of PM6 and PTO2 polymer-based OSCs, showing that low-offset systems display strong hybridization of the LE and CT states.[15] They further argued that hybridization accelerates recombination, thereby reducing the exciton-to-free charge conversion efficiency. In contrast, Jasiūnas *et al.* attributed the poorer charge generation efficiency in low offset blends to fast hole-back transfer.[16] Finally, Müller *et al.* conclude that a reduced IE offset strongly hinders CT splitting as compared to LE dissociation, the former being attributed as the main factor responsible for the resulting low charge generation.[17]

While different pictures were proposed about why low $\Delta$IE limits device performance and over which range this deteriorates the PCE, there is growing consensus that free charge generation in low-offset OSCs is assisted by the internal electric field. Several groups reported an effect of the internal electric field on the photovoltaic quantum efficiency ($EQE_{PV}$), the steady state photoluminescence (ssPL), transient PL (trPL) quenching, and the ground-state bleach signatures in transient absorption (TA) spectroscopy. These observations have in part been attributed to the field assisted dissociation of the CT states[17]–[20] and partially also to the field-assisted LE dissociation into CT states.[21]–[23] It is critical to note that some of these physical properties are a cumulation of multiple processes in the full device that can potentially have different bias-dependences. For instance, bias-dependent $EQE_{PV}$ spectra can be caused by the electric field-dependence of not just charge generation, but also charge extraction. Therefore, it is crucial to disentangle these processes. Bias-dependent PL quenching has been studied in fullerene-based OSCs as well as in NFA-based systems as a means to study dissociation efficiency of LE and CT states.[13], [15], [17], [20], [24] While it is true that a field-induced quenching of the ssPL intensity always indicates a depletion of the radiative LE states, other processes such as the repopulation of LE from non-dissociated CT states, possibly coupled to the reformation of CT states from free charges, are also influenced by the internal electric field. These additional processes may play a significant role in low-offset systems where the LE and CT populations are in kinetic equilibrium.[14], [16], [25]

In this work, we systematically explore the role of low driving force to free charge generation through a methodical assessment of generation and emission properties in a sample set of Y-series based OSCs. We present the results from blends of the NFAs Y5 and Y6 with two separate molecular weights of the polymer donor PM6 (Figure 1). By using PM6 with different molecular weights and two NFAs with different termination of the conjugated core (H- versus F-), we reduce the ΔIE in the blend from 350 meV to 210 meV, accompanied by a strong reduction of the device power conversion efficiencies (PCEs) from 15% to 1%. Poor PCEs are mainly caused by a pronounced field-dependence of free charge generation, probed *via* time-delayed collection field (TDCF) measurements. TA spectroscopy shows strongly diminished electro-absorption and polaron bands of the donor polaron in inefficient blends, accompanied by negligible formation of excited states (ground state bleach) of the donor upon selective NFA excitation. In addition, bias-dependent TA spectra show accelerated formation of the donor ground state bleach and electro-absorption bands in the presence of an external field in devices. We further demonstrate that the enhancement in free charge generation under an effective field is accompanied by a concomitant reduction in ssPL, which is dominated by the emission of the NFA LE.[11], [26] Most importantly, we are able to accurately reconstruct the field-dependent photoluminescence quantum efficiency (PLQY) from the PLQY of the neat acceptor for all of the studied systems, taking into account the free charge generation efficiency from TDCF and the LE reformation efficiency, obtained from the combination of electro- (EL) and photoluminescence. Based on these findings, we conclude that photocurrent losses by geminate CT recombination are of very minor importance. Moreover, the emission properties of the NFA LE in the blend are found to be little affected by the presence of the donor polymer, *i.e.* by hybridization with the CT state. With this knowledge at hand, we are able to reproduce the large effect that ΔIE has on the photogeneration efficiency.

## Results:

### a) Optoelectronic Properties

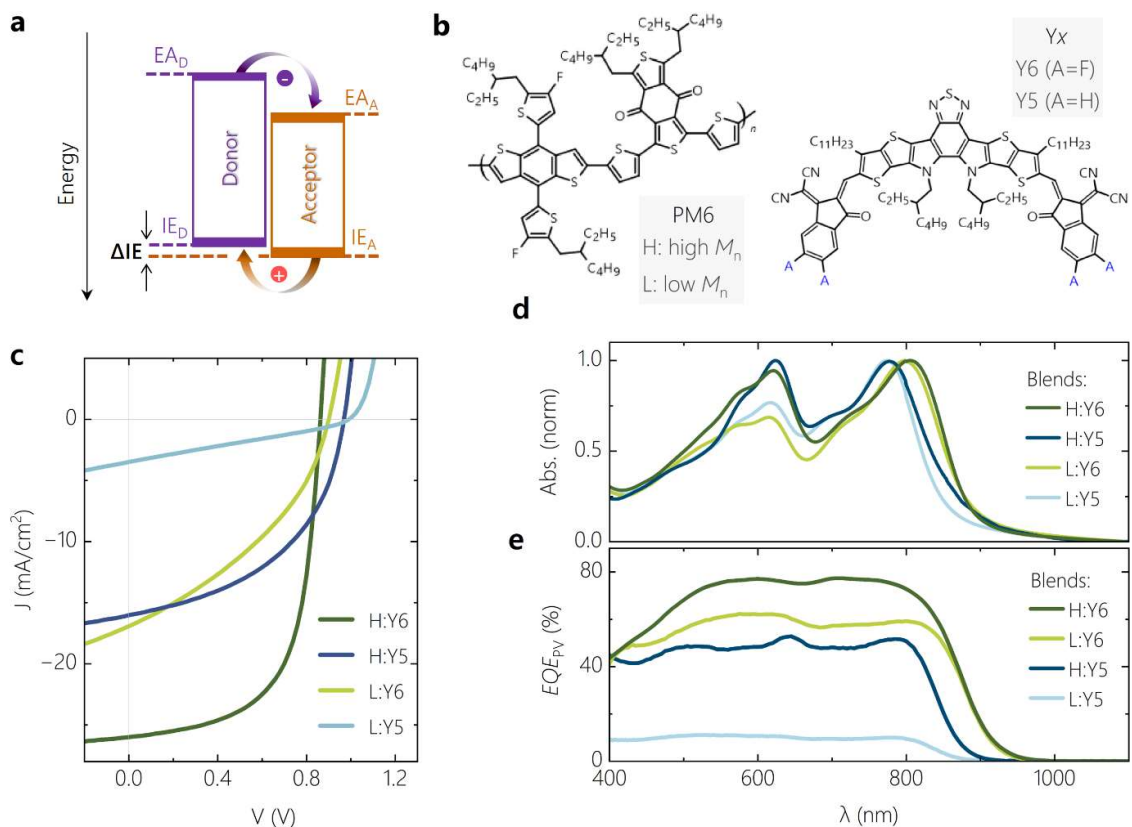

*Figure 1. Molecular structure, electronic, optical and photovoltaic properties of model systems. (a) Schematic representation of a D:A heterojunction with relevant electronic levels and possible charge transition routes. (b) Molecular structures of PM6, Y5 and Y6. (c) Current voltage (JV) characteristics measured under simulated AM1.5G illumination, depicting the performance range of the four OSC model systems. (d) Normalised absorbance of blend films of comparable thickness and (e) photovoltaic quantum efficiency ($EQE_{PV}$) of the four model systems measured under short-circuit conditions.*

**Figure 1b** shows the molecular structure of the polymer donor PM6 along with the two NFAs Y5 and Y6. Details on the synthesis and molecular weight of PM6 and full chemical names of the components are given in Supplementary Note 1. The two NFA molecules differ in that Y6 is fluorinated while Y5 is not. As a consequence, a ca. 0.1 eV smaller IE has been reported for Y5.[27] We combined these two NFAs with two batches of PM6 with different molecular weights to further fine-tune the $\Delta$IE: a relatively high number-average molar mass ($M_n$) of ca. 34 kDa, labelled as H or $H_{PM6}$, and a low $M_n$ of ca. 3.5 kDa, denoted as L or $L_{PM6}$.[28] As shown in previous works, the $L_{PM6}$ polymer aggregates less in neat and blend films, suggesting an increase in IE and a reduced $\Delta$IE of the blend.[28], [29]

The current-voltage (JV) characteristics of regular devices with comparable thicknesses of the photoactive layer (100-110 nm) are shown in **Figure 1c** (see **SI Figure S1** for the statistics of the photovoltaic parameters). While the H:Y6 blend exhibits a typical PCE of up to 15%, the device performance drops considerably upon replacing Y6 by Y5. This is because the reduction in $J_{SC}$ and *FF* overcompensates the increase in $V_{OC}$. Our earlier studies suggested inefficient and field-dependent exciton dissociation as the primary reason for the poor performance of the PM6:Y5 blend, related to a smaller $\Delta$IE.[11], [26], [30] Replacing $H_{PM6}$ by $L_{PM6}$ changes the device parameters in a similar way, suggesting a reduced ΔIE as the main cause for the poor performance also for this blend, herein arising from reduced polymer aggregation as described earlier. This varies the PCE by more than a factor of ten, from on average 15% for H:Y6 down to 1% for L:Y5, while keeping the chemical structure of the conjugated backbone of the constituents nearly the same. We note here that a strong decrease in performance was reported by Karki *et al.* when blending Y6 with a ca. 1:1 mixture of high and a low $M_n$ PM6, which was attributed mainly to a different bulk and interface morphology.[29] We will show later that the reduction in PCE is not primarily caused by differences in charge extraction. Evidence for the different aggregation properties of the two molecular weight batches, in neat layers and in the blends, comes from the absorption spectra in **SI Figure S2** and **Figure 1d**, respectively. While the shape and spectral position of NFA absorption in the blend is nearly unaffected by the $M_n$ of PM6, replacing of $H_{PM6}$ with $L_{PM6}$ leads to an overall reduction in the polymer peak absorption strength, but also of the 0-0:0-1 peak ratio. Based on the results of absorption spectroscopy on neat PM6,[28], [31] these spectral changes are interpreted to originate from weaker polymer chain aggregation. There is also a notable influence of the choice of the NFA on the polymer absorption spectra. For both $H_{PM6}$ and $L_{PM6}$, polymer aggregation seems to be less pronounced in blends with Y6. This can be explained by the stronger tendency of Y6 to form extended aggregates by itself, which competes with the formation of PM6 aggregates in the drying blend film. Previous works showed a higher tendency of Y6 to aggregate compared to Y5, which was assigned to stronger intermolecular charge transfer (ICT).[32]–[34] The corresponding *EQE*$_{PV}$ spectra are shown in **Figure 1e**. Replacing Y6 with Y5 and/or $H_{PM6}$ with $L_{PM6}$ reduces *EQE*$_{PV}$ over the entire spectral range but has little effect on the shape of the spectra. *EQE*$_{PV}$ spectra generally differ from the absorption spectra because photons which are reflected at the cathode also contribute to photocurrent generation. Obviously, the reduced photocurrent generation in our OSCs with Y5 or $L_{PM6}$ does not originate from effects related to individual material properties such as very short exciton diffusion lengths in one of the blend components. We propose the free charge generation process in such low-offset OSC systems is dictated by the hole-transfer process even when the excitons are initially generated in the donor, in agreement to results from previous studies.[12]

## b) Blend Energetics

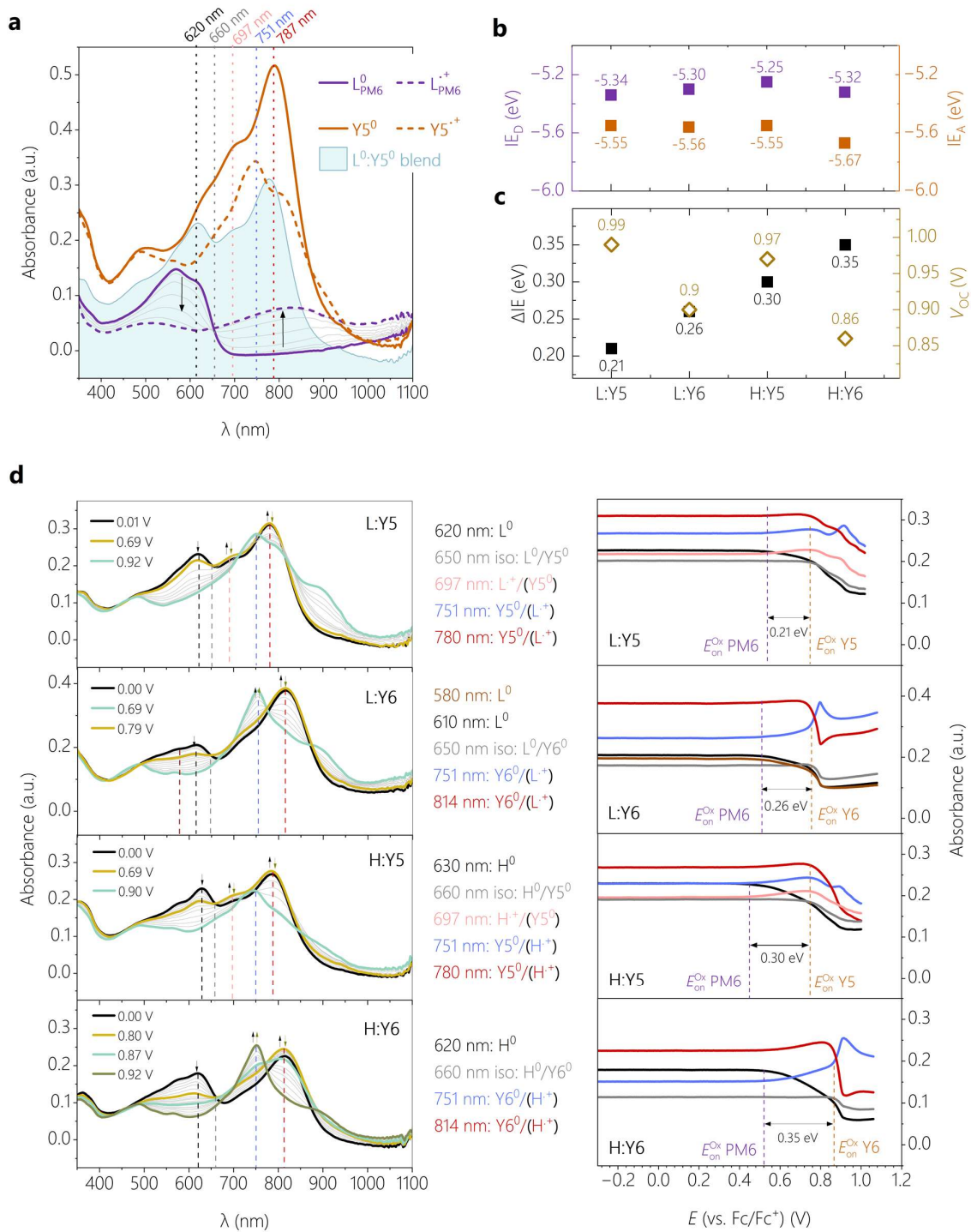

***Figure 2. Blend energetics.*** *(a) Absorption spectra recorded during cyclic voltammetry of the neat films of L$_{PM6}$ and Y5, and the blend of L$_{PM6}$ and Y5 in the neutral form (at 0.00 V vs Fc/Fc$^+$ for neat and blend) and the radical cation form of neat L$_{PM6}$ (0.97 V vs Fc/Fc$^+$) and Y5 (0.92 V vs Fc/Fc$^+$). The vertical dashed lines are guides-to-the-eye to highlight characteristic bands of the different species. (b) Ionization energies (IE) of donor and acceptor measured in the blends. (c) Comparison of open circuit potential V$_{OC}$ with the IE offsets ΔIE as determined from the data in (b). (d) Absorption spectra (left) for selected potentials recorded during oxidation, and the corresponding peak trends (right) for the different blends L$_{PM6}$:Y5, L$_{PM6}$:Y6, H$_{PM6}$:Y5 and H$_{PM6}$:Y6 (top to down).*

*In-situ* spectroelectrochemistry (SEC), i.e. coupling cyclic voltammetry with in-situ UV-vis-NIR spectroscopy, is used to determine the IEs of the four blends with identical preparation as used for the solar cell characterization with ITO as the substrate. This technique has proven to be useful in terms of mapping the spectral evolution during charging in electrochemical experiments and determining the oxidation and reduction onsets from the spectral onsets of neutral and first charged species (polaron).[35] In particular for blend films, there lies a problem of overlapping charged states which makes it impossible to extract the onsets from the cyclic voltammograms alone. In an earlier study we have shown the advantages of the spectral onset determination for the system PM6:Y6 with different molecular weights, spin-coated from other solvents.[35] A general finding was that PM6 has a lower oxidation onset than Y6, i.e. higher IE. Here, we focus on the oxidation of the blend films which involves the potential scan to positive potentials. From the oxidation onsets, the IEs and ΔIEs can be calculated which will be later used for a comparison with the open circuit potentials $V_{oc}$. The oxidation of neat films of the single compounds (identical preparation as for blends) was done in parallel and is used to help identify characteristic absorption bands in the blends.

In **Figure 2a**, the characteristic spectra of films of neat $L_{PM6}$, neat Y5 and the blend $L_{PM6}$:Y5 are shown for the neutral state. The corresponding CVs are summarized in the supporting information **SI Figure S3**. During a positive potential sweep, the neutral absorption bands of $L_{PM6}$ at 570 nm and 620 nm decrease and a broad bathochromically shifted band is appearing with its peak at 820 nm, indicating the radical cation form $L_{PM6}^{\cdot+}$. One can also clearly identify a characteristic isosbestic point at 660 nm. For neat Y5, the spectral response during cyclic voltammetry is a decrease of the 787 nm band of the neutral molecule and the appearance of a new hypsochromically shifted band maximum at 751 nm, which is assigned to the formation of the radical cation form Y5$^{\cdot+}$. The neutral blend spectrum $L_{PM6}$:Y5 resembles the neutral spectra of the single components, with slight shifts which can be explained by the superposition of the spectra. The maximum around 620 nm in the blend spectrum can be attributed to $L_{PM6}$, whereas the maximum at 780 nm to the Y5 species. Following the same principle, the main PM6 and the main Yx absorption bands can be assigned in all blends **(SI Figure S4)**. For the Y6 systems, the maximum of Y6 is bathochromically shifted to 814 nm with respect to Y5.

In **Figure 2d (left)**, the spectra of the four blends during oxidation are shown in the potential range between 0.00 and ∼ 1.00 V vs Fc/Fc$^+$, the intensity evolution of characteristic absorption bands is highlighted in Figure 2d (right). The corresponding oxidation cycles can be found in **SI Figure S5**. For $L_{PM6}$:Y5**,** increasing the potential from 0.00 V (black spectrum) to 0.69 V (dark yellow spectrum) goes along with a decrease of the intensity of the absorption maximum at 620 nm which is characteristic for the continuous oxidation of the neutral $L_{PM6}^{0}$. Parallel to this, the increase of absorption at 697, 751 and 780 nm indicates the formation of the radical cation $L_{PM6}^{\cdot+}$. Consistent with the spectra evolution of neat $L_{PM6}$ an isosbestic point at 650 nm can be observed in this potential range. The peak evolutions in **Figure 2d (right)** can be used to determine the oxidation onsets via the tangent method.[35] This method involves the transfer of the potential onsets to the Fermi scale with a correction factor of -4.8 eV. Hereby, the bands at 620 nm and 697 nm are used to determine the oxidation onset $E_{on}^{ox}$($L_{PM6}$) of PM6, which gives 0.54 V equivalent to IE$_D$ = -5.34 eV. Increasing the potential from 0.69 V to ∼ 1.00 V, the band at 780 nm is decreasing and the band at 751 nm is appearing, showing the oxidation of neutral Y5$^{0}$ to the radical cation Y5$^{\cdot+}$. Unfortunately, the oxidation of Y5 goes along with an overall decrease of the absorption between 600 and 900 nm, which aggravates the determination of the exact oxidation onset from these spectra. We, therefore, base the determination of the oxidation onset $E_{on}^{ox}$(Y5) of Y5 on the absorption at the isosbestic point of PM6 in the blend at 650 nm, which yields a value of 0.75 V or an IE$_A$ of -5.55 eV.

The evaluation of the blend $H_{PM6}$:Y5 follows qualitatively the same procedure as for the $L_{PM6}$:Y5. Hereby, the oxidation onset of $E_{on}^{ox}(H_{PM6})$ is determined at 0.45 V, equivalent an $IE_D$ of -5.25 eV. The oxidation onset of $E_{on}^{ox}(Y5)$ is again determined from the absorption at the isosbestic point of PM6, yielding 0.75 V or an $IE_A$ of -5.55 eV. The IEs of the Y6-based blends have been analyzed the same way. Here, we additionally benefit from the more pronounced increase of the absorption at 751 nm, assigned to the radical anion of Y6. Note that different $L_{PM6}$:Y6 samples revealed slightly different spectral shapes but also different oxidation onsets, while the value of ΔIE was reproducible at $0.30 \pm 0.02\ V$. Probably this can be explained by the low molecular weight of this polymer and the lower aggregation tendency, which might lead to dissolution of the blends during the electrochemical experiment. Therefore, the absolute values of $IE_A$ and $IE_D$ of $L_{PM6}$:Y6 should be interpreted with caution. Here, we show and analyzed the SEC spectra of a blend which revealed the clearest PM6 oxidation onset in the CV.

**Figure 2b** plots the ionization energies of the four blends. With the exemption of the $L_{PM6}$:Y6, the trends follow the expectations from the blend absorption as discussed earlier. For the Y5-based blends, the ca. 0.1 eV smaller IE of $H_{PM6}$ compared to $L_{PM6}$ is consistent with a stronger aggregation of the high molecular weight PM6 in the blend with Y5. In contrast, replacing Y5 by Y6 in the blend with $H_{PM6}$ increases the $IE_D$, bringing it back to the IE of $L_{PM6}$ in the blend with Y5. Regarding $IE_A$ we find similar values of the Y5 NFA in two blends, which agrees with the very similar Y5 absorption in both blends. The IE of Y6 in $H_{PM6}$ is ca. 0.1 eV higher than of Y5, which is consistent with the reported down-shift of the HOMO upon fluorination.[27] The $L_{PM6}$:Y6 shows overall too small ionization energies, which is a direct outcome of the SEC described above.

The resulting ΔIE is compared to the device $V_{OC}$ in Figure 2c. The highest performing $H_{PM6}$:Y6 blend, with a PCE of over 15 % and a $V_{oc}$ of 0.86 V, has the largest IE offset of 0.35 eV while the poorest performing blend, $L_{PM6}$:Y5 with an PCE of 1% and a $V_{oc}$ of 0.99 V, has the smallest IE offset of 0.21 eV. Here we remind the reader that the correct determination of ΔIE in the PM6:Y6 blend is among the hottest debated topics of research on the PM6:Y6 blend.[36] In the past, these energies were mostly derived from CV and UPS on neat layers, with large differences in the final results. For example, the CV data in the first PM6:Y6 paper[37] yielded an ΔIE of less than 0.1 eV while UPS data from different sources yielded ΔIE>0.5 eV.[27] However, a ΔIE of < 0.1 eV is too small to split the Y6 excitons, which have a binding energy of ca. 0.3 V.[35] On the other hand, with the well-established fundamental gap of 1.7 eV for Y6, ΔIE of ca. 0.5 eV translates into an energy of the charge separated states ($E_{CS}$) of less than 1.2 V, which is rather small in light of a $V_{oc}$ of 0.83 V. A ΔIE of ca. 0.3-0.34 eV as reported here and in our previous work for H:Y6[35], should be sufficiently large to dissociate the Y6 exciton while providing a high enough $E_{CS}$ of 1.4 eV to explain the blend $V_{OC}$. Regarding our other blends, the increase in $V_{OC}$ relative to H:Y6, is qualitatively in agreement with the trend in ΔIE when taking into account that Y5 has a ca. 30 meV larger HOMO-LUMO gap than Y6.[27] As outlined earlier, ΔIE provides an only rough estimate of the energy of the CT state, the latter being relevant for the value of $V_{OC}$ in most organic solar cells.[38]

## c) Anticorrelation of Field-dependent Free Charge Generation and Emission

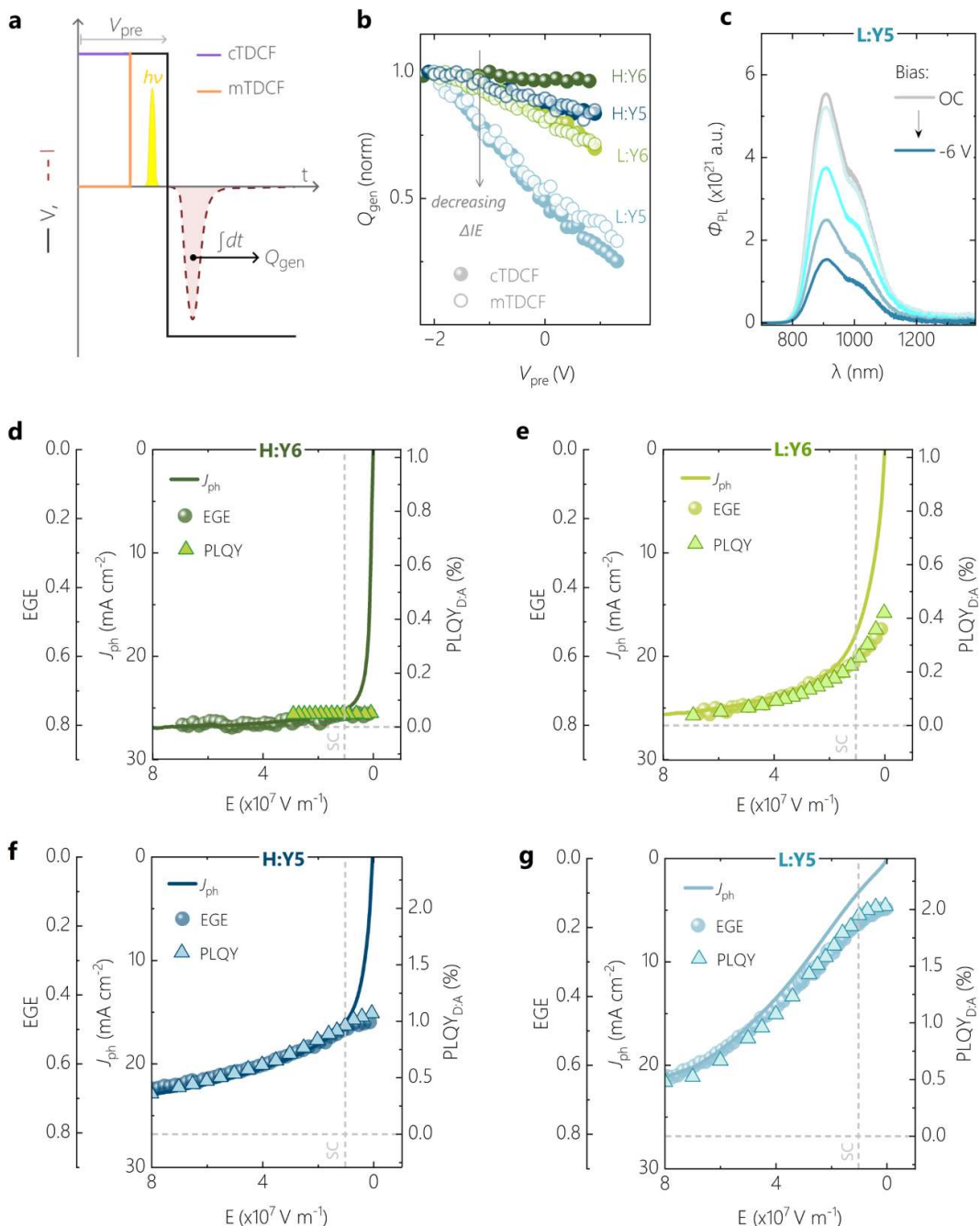

**Figure 3. Ubiquitous anticorrelation of bias-dependent LE emission and free charge generation.** (a) Schematic illustration of classical and modified time-delayed collection field (cTDCF and mTDCF) techniques during one extraction cycle. (b) Normalised comparison of the photogenerated charge $Q_{gen}$ as a function of applied pre-bias $V_{pre}$, measured using cTDCF (filled spheres) and mTDCF (hollow circles), for all systems except H:Y6. (c) An example of steady state photoluminescence (ssPL) spectra, quenched with applied reverse bias, shown for the L:Y5 system. (d) Overlay of field-dependent photocurrent density ($J_{ph}$), external free charge generation efficiency (EGE) from TDCF, and $PLQY_{D:A}$ (reconstructed from ssPL photon flux) for the H:Y6 OSC, showing very good anticorrelation of generation and emission. The same is plotted for L:Y6, H:Y5 and L:Y5 in (e), (f), and (g), respectively. Horizontal dashed lines denote zero emission from the NFA in the blend, and vertical dashed lines denote the electric field at short-circuit (SC) condition.

In our previous studies on a blend of high $M_n$ PM6 with Y5, we showed that an anticorrelation exists between free charge generation and the intensity of ssPL.[26] This provided us with evidence that the bias-dependence of $J_{ph}$ in this blend is largely determined by the field-dependence of the dissociation of the Y5 LE exciton. In the following, we will build on this by demonstrating quantitative relations between the bias-dependence of $J_{ph}$, the external charge generation efficiency (EGE) and of the photoluminescence quantum efficiency of the blend ($PLQY_{D:A}$). As in Ref 20, our method of choice to measure the efficiency of exciton to free charge conversion is TDCF, see[39], [40] for details. A low laser fluence of ca. 60 nJ/cm$^2$ ($\lambda$=532 nm, pulse width=ca. 6 ns) was used to excite the device held at a pre-bias ($V_{pre}$), followed by immediate collection of photogenerated free charge carriers using a high reverse collection bias (see **Figure 3a**). In our current setup, a fast slew rate of 2 ns of the voltage source enables fast extraction of the photogenerated free charge ($Q_{gen}$). While these experimental conditions should minimize losses due to non-geminate recombination (NGR) among photogenerated charge carriers, recombination between photogenerated and dark injected charge carriers might be an issue, especially at positive $V_{pre}$.[41] To check whether this loss process is an issue in our samples, we performed additional measurements with our newly developed modified-TDCF, or mTDCF, technique.[26] In contrast to the classical method (cTDCF), the pre-bias is herein applied for only 25 ns which is three times the RC time of our sample, so that the density of dark-injected charge carriers available in the bulk for NGR should be largely minimized. **Figure 3b** shows the $Q_{gen}$ from cTDCF and mTDCF for the present four systems as a function of the pre-bias voltage, normalized at $V_{pre}$ =-2 V. Both techniques demonstrate the same field-dependence of free charge generation, though our measurements were performed to above $V_{pre}=V_{OC}$. It is only for the highly inefficient L:Y5 device that we measure up to 10% less free charge from cTDCF compared to mTDCF around $V_{OC}$ conditions. While this indicates extraction losses due to recombination with dark charge in the positive pre-bias range, the effect is too small to account for the pronounced field dependence of free charge generation derived from TDCF for this system. The results in **Figure 3b** therefore highlight a clear correlation of progressive bias-dependent free charge generation with a diminishing ΔIE in the D:A blends.

Knowing $Q_{gen}$ and the excitation fluence of the incident photons, we calculate the bias-dependence of the external free charge generation efficiency, or EGE. Then, the so-obtained EGE(V) is compared to $J_{ph}$(V) as a function of the applied bias. As shown in **Figure 3d-g**, EGE(V) overlaps almost perfectly with $J_{ph}$(V) for all four systems when scaling the two properties such that an EGE of 80 % corresponds to a $J_{ph}$ of 27 mA/cm$^2$, i.e. of the high performing H:Y6 blend (see **SI Figure S6** for a condensed representation of the data). It has been shown that photon losses due to reflection and parasitic absorption account for a ca. 20% loss of maximum photogenerated free charge.[42] Deviations between EGE(V) and $J_{ph}$(V) are present mainly at low electric fields, arising from NGR which additionally reduces $J_{ph}$. These non-geminate losses become more pronounced with decreasing ΔIE. It was shown in recent studies that low-offset D:A blends exhibit a larger NGR coefficient, though the reason for this correlation is still unknown.[11] We conclude that $J_{ph}$(V) is governed by the bias-dependence of free charge generation over a wide voltage range with only small contributions of NGR. Notably, even the low performing L:Y5 blend achieves an EGE larger than 60 % at a -7 V reverse bias, more than three times the value at the short-circuit condition.

At this point, a question arises whether the bias-dependence of EGE of these systems, measured by predominantly exciting the donor at $\lambda$=532 nm, is fully representative of the generation properties of each D:A blend over a wide spectral range. Consider the scenario where ΔEA is larger than ΔIE, as is the case in our systems. If photogenerated excitons in the donor would, for instance, dissociate at a D:A interface prior to resonant energy transfer to the NFA, then electron transfer from donor excitons is expected to be more efficient and hence less bias-dependent than hole transfer from NFA excitons. To address this, we recorded $EQE_{PV}$ for different biases on the two D:A systems with strongest bias-dependent EGE, i.e. L:Y6 and L:Y5. As shown in **SI Figure S7a**, an increasing negative bias expectedly reduces the $EQE_{PV}$ over the entire spectral range for both systems. Inspection of the normalised $EQE_{PV}(V)$ of L:Y6 in **SI Figure S7b** shows the shape of the spectra to be totally unaffected by the bias, while the normalised $EQE_{PV}(V)$ of L:Y5 exhibits a marginally enhanced bias-effect when exciting the Y5 acceptor than when exciting the $L_{PM6}$ donor. TDCF experiments with selective excitation of the Y5 NFA at λ=800 nm prove that the free charge generation is indeed marginally more bias-dependent, but that this effect is small compared to selective donor excitation at λ=532 nm (a difference of ca. 5% at $V_{pre}=V_{OC}$, see **SI Figure S8c and S8d**). This provides further evidence that the hole transfer pathway is indeed the critical step in the photogeneration process for equally all blend systems, which we later corroborate with transient absorption data in sub-ps timescales.

We next measured the bias-dependence of ssPL spectra at a fixed 520 nm excitation and at an intensity so as to produce the same short-circuit current as under AM1.5 illumination. These measurements were taken on the very same device structures as TDCF and it was ensured that the excitation spot is limited only to the device area. For all blends, the spectral position and shape of the PL differs only little from that of the neat acceptor dispersed in a polystyrene matrix (PS:NFA) with same concentration as in the D:A blends (denoted as PS:Yx in the following), see **SI Figure S9**. Such spectral differences are likely due to microcavity effects in the full device, as well as minor differences in the NFA aggregation properties.[43], [44] The overlapping ssPL from D:A and PS:NFA shows that the PL in blends with the donor is governed entirely by the radiative decay of the respective NFA LE state with very little contribution from radiative CT decay. For the blends that exhibit a bias-dependent $J_{ph}$, we observe a strong reduction of the PL photon flux ($\Phi_{PL}$) with increasingly negative reverse bias, as can be seen for L:Y5 in **Figure 3c** and for L:Y6 in **SI Figure S9b** (H:Y5 has been previously reported[26]), while the spectral shape is not affected by the bias (**SI Figure S9a and S9c**). If radiative recombination of the CT state were to contribute to the ssPL spectrum, its contribution would decrease rapidly with higher negative bias because charge extraction hinders the reformation of CT states via NGR. In such a case, spectral changes would be expected.[39] In addition, we measured the EL spectrum at a voltage that produces the same recombination current as in the ssPL experiment. We find that the shape of the EL spectrum is very similar to the device ssPL, while its intensity is significantly lower (see **SI Figure S9b and S9d**). This confirms our view that all emission from these devices occurs via the local LE excitons.

In **Figure 3d-g**, we correlate $\Phi_{PL}$ to $J_{ph}$ and EGE as a function of the effective electric field $E = (V_{bi} - V)/d$, where $V_{bi}$ is the built-in voltage of the respective OSC. To do so, the bias-dependent PL peak intensities ($\Phi_{PL,max}$) were first converted into a PL quantum yield ($PLQY_{D:A}(E)$), extrapolated from the measured PLQY of the device at $V_{OC}$ using

$$PLQY_{D:A}(E) = \frac{\phi_{PL,max}(E)}{\phi_{PL,max}(E\ at\ V_{OC})} \cdot PLQY_{D:A}(V_{OC}) \qquad (1)$$

Here, $PLQY_{D:A,}(V_{OC})$ was measured on a full device structure in an Ulbricht sphere under 1 sun equivalent illumination at open-circuit conditions. Herein it was ensured that the excitation laser beam in the Ulbricht sphere only illuminated the active area of the complete OSC device stack.

To correlate $PLQY_{D:A}(E)$ with EGE(E) and $J_{ph}(E)$, we first aligned $PLQY_{D:A}=0$ to EGE=80%; the condition where we concluded that all photogenerated excitons are converted into free carriers. We were then able to establish a nearly perfect anticorrelation between the field-dependences of $PLQY_{D:A}$ and EGE for all four blends when we aligned EGE=0% to the PLQY of the corresponding inert PS:Yx blend, i.e. $PLQY_{PS:A}$ – where all recombination proceeds through the decay of the Yx excitons. This quantitative agreement leads us to the conclusion that the field-enhanced free charge generation comes directly at the cost of quenched excitonic emission. In other words, geminate exciton recombination is the main (if not only) decay channel competing with free charge generation. Note that the PLQY of the PS:Yx devices are field-independent (see **SI Figure S10**), which means that the internal electric field enhances LE dissociation only at the D:A heterojunction but not within the NFA domains.

### d) Charge Generation Dynamics in Transient Absorption Spectroscopy

We confirm these conclusions with measurements of the early time dynamics of the photoexcited species using TA spectroscopy. Given the fact that the ssPL is dominated by the NFA emission, we firstly performed TA measurements by selectively exciting the NFA at λ=720 nm, with a low pump fluence of ≈2 μJ/cm². TA spectra were recorded while probing in the visible (500-680 nm) and near infrared (NIR) regions (680 -1350 nm) of the electromagnetic spectrum, resolved in time delays between 0.1 ps and 7 ns. As a benchmark, the optical transitions in pure NFAs, probed in films of PS:NFA, are overlaid on the TA spectra of different D:A blends.

**Figure 4a** shows the visible and NIR TA spectra for H:Y6 at different time delays between the incident pump and probe pulses. At very short time delays (0.1-0.3 ps), we observe a negative feature due to the ground state bleach (GSB) of Y6 at λ>600 nm, along with a substantial GSB contribution from $H_{PM6}$ between 580-600 nm in the visible region.[45], [46] The presence of $H_{PM6}$ GSB upon selective pumping of Y6 provides confirmation of hole transfer at ultrafast time scales (<0.3 ps). A similar observation is made from the NIR region, wherein the formation of new positive bands between 700-750 nm is attributed to the electro-absorption (EA) feature of $H_{PM6}$, arising from the Stark-shift of the $H_{PM6}$ absorption due to the electric field caused by charge separated species.[45], [47] The EA feature is accompanied by spectral red shifts and broadening of the positive photoinduced absorption band (PIA), as referenced against the PIA feature of pure Y6 (peak at 925 nm). This is attributed to the growth of a polaron absorption band centred at 980 nm due to overlapping contribution from hole[29] and electron[36], [48] absorption bands of PM6 and Y6, respectively. With progression of the delay, both the $H_{PM6}$ GSB and the EA features rise and complete their evolutions at ≈100 ps (see **SI Figure S11c**). It is important to note that even with the initial sub-picosecond rise, the complete hole transfer process takes about 100 ps to reach its peak due to the time required for excitons formed within Y6 domains to diffuse to the interface, as has been previously reported for Y6 and several other NFAs.[49] In this context, we acknowledge recent TA work suggesting that free charge formation in Y6-based OSCs occurs mainly via the dissociation of intermolecular (intramoiety) charge transfer states (often denoted as xCT), especially at longer delay times.[46], [47] On the other hand, analysis of the *EQE*$_{PV}$ spectra of CuSCN/NFA bilayer devices with different NFA layer thickness [50] and of the intensity dependence of pulsed PL measurements on neat NFA layers[51] revealed similar values for the exciton diffusion lengths. This suggests that the xCT and the LE are in kinetic equilibrium (at least at room temperature) through, for example, thermal re-excitation of the LE state [47] and/or through strong electronic coupling .[52] The TA measurement for the L:Y6 blend with smaller ΔIE was performed under the same conditions for a clear comparison, see **SI Figure S11a** for the visible and NIR region TA spectra. Unsurprisingly, the spectral signatures of $H_{PM6}$ GSB, EA and the polaron absorption closely match those of H:Y6. However, compared to H:Y6, L:Y6 exhibited a muted rise in both $H_{PM6}$ GSB and EA features. The comparatively slow rise component of these features in L:Y6 can be observed in both the initial sub-picosecond regime as well as the generation timescales attributed to the exciton diffusion-dependent component (see **SI Figure S11c**). **SI Figure S11b** further shows the comparison of the negative GSB dynamics for H:Y6 and L:Y6 blends, wherein the H:Y6 (L:Y6) reaches its peak at ≈100 ps (≈200-300 ps). As a consequence, exciton recombination competes efficiently with charge generation characteristics in L:Y6, in full support of our findings from TDCF and ssPL. **SI Figure S11c** shows the comparison of PM6 GSB and the EA feature for H:Y6 and L:Y6 blends. Clearly, the EA feature is found to closely mirror the rise in PM6 GSB, suggesting instantaneous dissociation of and almost no accumulation in the population for CT states.

Next, we move our focus to the Y5 acceptor blends. **SI Figure S12a** shows the visible and NIR region TA spectra for H:Y5 at a pump fluence of 2 μJ/cm$^2$ . The GSB of Y5, peaking at 640 nm, shows a minor rise in intensity at early time delays (<5 ps) due to hole transfer, resulting in an overlapping contribution from H$_{PM6}$ at this wavelength and this is followed by a strong decay of Y5 excitons as the time progresses. At 50 ps, a GSB feature arises around 580 nm – which is selectively attributed to H$_{PM6}$ GSB – following the decay of the overlapping positive feature of PIA from the Y5 exciton. Similarly, in the NIR region, the positive exciton PIA band at 900 nm Y5 shows a minor red shift and broadening due to the overlapping hole absorption of PM6 in the early time scales, and this is followed by a consistent decay due to inefficient dissociation of Y5 excitons. This is in clear contrast to the H:Y6 blend, wherein the strong contribution from H$_{PM6}$ GSB is visible from the very early time delays due to a more efficient hole transfer. In most contrast to H:Y6, **Figure 4b** shows the TA results for L:Y5, the most inefficient blend with the smallest ΔIE, under the same pump fluence. Both visible and NIR TA spectra are dominated by excitonic decay of GSB and PIA features as referenced against the pure Y5. The negligible hole transfer in the blend is clear from the feeble generation of L$_{PM6}$ GSB at the 580-590 nm wavelength window. To get a meaningful comparison of hole transfer and free charge generation efficiency between H:Y5 and L:Y5 blends, we plot the GSB dynamics at 630-640 nm, corresponding to overlapping GSB of PM6 and Y5, see **SI Figure S12b.** The comparison of dynamics at early time scales shows an obvious rise in GSB intensity in H:Y5 which is absent in L:Y5. **SI Figure S12c** shows the longer time dynamics normalised to the initial intensity of the two Y5-based blends along with the normalised GSB decay of PS:Y5. H:Y5 clearly shows a higher intensity at longer time due to the formation of longer-lived free charge carriers, while the dynamics of L:Y5 are closely aligned with that of PS:Y5. These results closely follow the trend shown previously in TDCF generation at $V_{OC}$ and signify the inefficiency in exciton dissociation in the low offset systems.

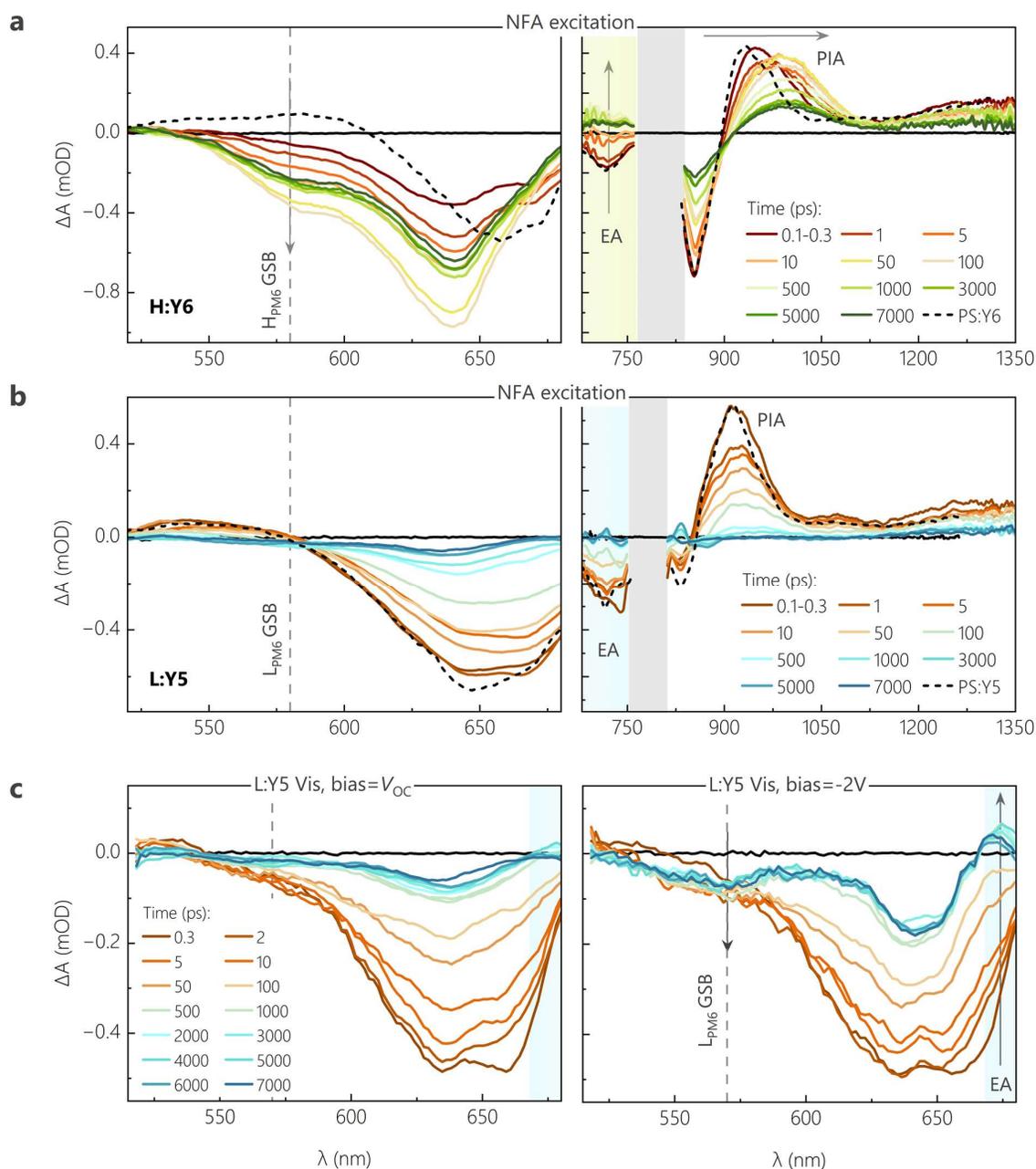

**Figure 4. Spectroscopic investigation (open-circuit and biased) of dynamics of free charge generation reflects field-dependent performance of model NFA blends.** *Transient absorption spectra in the visible and infrared region, probed on films of the (a) H:Y6 blend and (b) L:Y5 blend, excited with a 1.77eV laser pulse of 2 μJ/cm² fluence for preferential NFA excitation. The grey shaded area denotes the region of optical excitation of the probe beam. The black dotted lines denote the optical transitions of the respective PS:NFA films at 1 ps in order to benchmark the transient spectra of the blends. (c) Bias-dependent transient absorption spectra in the visible region probed on an L:Y5 device, when biased externally with V_OC and -2 V. The semi-transparent sample was excited with a 1.77eV laser pulse of 6 μJ/cm² fluence for selective NFA excitation.*

To confirm and corroborate the observation of field-dependent exciton dissociation in earlier sections, we performed bias-dependent TA measurements on H:Y6 and L:Y5 blend systems due to their very different field-dependent free charge generation characteristics shown in TDCF measurements.

Measurements were performed on semi-transparent devices with the same device structure as that used for photovoltaic and TDCF studies, but with the thickness of silver electrode reduced from 100 to 15 nm. The biased TA measurements were performed in the visible region of the probe to avoid artefacts from longer probe wavelengths due to the semi-transparent contact.[53] To achieve sufficient signal to noise ratio, a pump fluence of 6 $\mu$J/cm$^2$ was used for both devices. The biased-TA response for H:Y6 device under $V_{OC}$ conditions and at -2 V, shown in **SI Figure S13a**, was found to be comparable and also similar to those recorded on bare films. The GSB feature around 580 nm (selective to PM6) shows a fast sub-picosecond rise along with a relatively slow rise, reaching its peak after ca.70-80 ps. **SI Figure S13b and S13c** shows the TA dynamics of the different GSB bands of H:Y6 devices at 580 and 640 nm, respectively, under different bias conditions. Clearly, the presence of an external field does not alter the free charge generation characteristics of this system. We note that the dynamics at 580 nm in biased TA of devices are slightly different for those seen in bare films and can be attributed to the relatively higher fluences used. The field-dependence of generation is, however, clearly observable in the TA spectra for L:Y5 devices at $V_{OC}$ conditions and -2 V, shown in **Figure 4c**. At $V_{OC}$, the TA spectra show a continuous decay of the Y5 GSB due to the decay of Y5 excited states and almost no contribution from the $L_{PM6}$ GSB at lower wavelengths (570-580 nm), suggesting feeble hole transfer in this blend system at $V_{OC}$ conditions. This is consistent with the TA results on bare L:Y5 films in **Figure 4b**. Under a bias of -2 V, the GSB decay of Y5 is contrastingly subdued with relatively higher intensity of the GSB at longer time delays (>500 ps), suggesting a higher contribution from long-lived free charge carriers. Further, we notice a clear rise in GSB contribution at lower wavelengths (520-570 nm band) in early time scales up to 5 ps, which can be safely attributed to the $L_{PM6}$ GSB as a result of the hole transfer process. This is also accompanied by a negative absorption band at 670 nm which is attributed to the EA feature arising from the generation of free charge carriers. The clear rise in $L_{PM6}$ GSB with an external bias, along with the concomitant rise in EA, confirms that the poor performance of this blend system indeed originates from the inefficient dissociation of LE states to CT.

To complete the TA study, we addressed the role of electron transfer. To this end, we performed TA experiments on the poorest performing blend system, L:Y5, with preferential excitation of the donor at 400 nm. At this excitation wavelength, $L_{PM6}$ is found to contribute ca. 83% of the total blend absorption as benchmarked by comparing the L:Y5 blend with PS:Y5 films. **Figure 5a** shows the resultant visible and NIR TA results for the L:Y5 blend, using a pump fluence of 3 $\mu$J/cm$^2$ (to yield a similar exciton density as in selective NFA excitation measurements). At 0.1 ps, the GSB of L:Y5 is found to be dominated by $L_{PM6}$ with a shoulder at 575nm, as shown by the overlapping GSB of the neat $L_{PM6}$ in **Figure 5a**. As time progresses, this feature is rapidly quenched along with the simultaneous rise of a GSB feature between 650-680 nm attributed to the GSB of Y5. This is further accompanied by the rapid reduction in the feature at around 1150 nm assigned to the PIA of LE to higher energetic states of $L_{PM6}$, as seen in **Figure 5a**. **Figure 5b** shows the comparison between the kinetic traces of $L_{PM6}$ PIA at 1150 and the $L_{PM6}$ GSB feature at 575 nm, wherein both demonstrate almost complete overlap in the early sub-picosecond timescales. The simultaneous reduction in the exciton PIA and the GSB of $L_{PM6}$ in the early timescale suggests that PM6 LE states rapidly undergo Förster resonance energy transfer to yield Y5 LE states, and hence the overall charge generation step is limited by poor hole transfer.

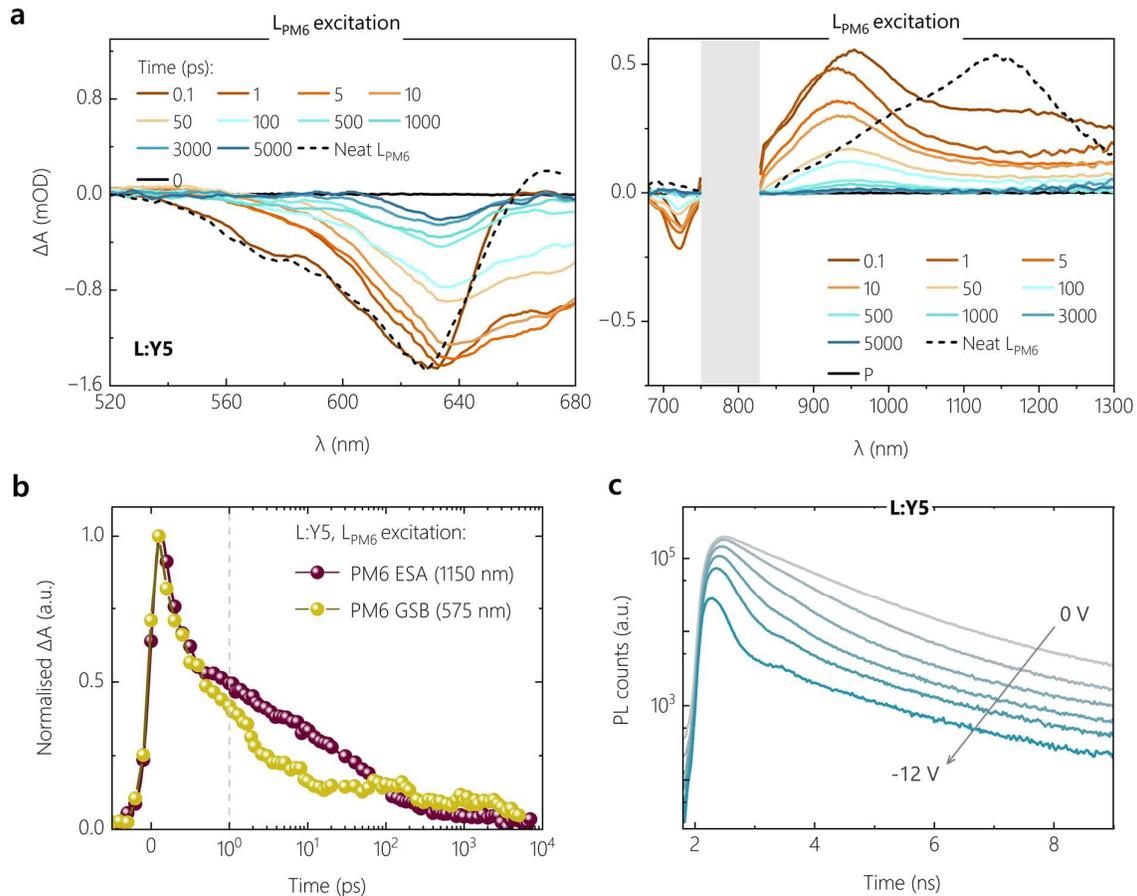

*Figure 5. Preferential donor excitation TAS, and transient photoluminescence trPL.* *Transient absorption spectra of L:Y5 at pump excitation wavelength of 400 nm in the visible (a) and infrared region (b) at a pump excitation wavelength of 400 nm. The optical transitions of $L_{PM6}$ at 1 ps are overlaid as a dotted line. The grey shaded area in the infrared region denotes the region of optical excitation of the probe beam. (c) Comparison of kinetic traces of the exciton PIA and GSB features of $L_{PM6}$, demonstrating efficient Förster resonance energy transfer as the dominant process. (d) trPL decays for the L:Y5 device as a function of applied reverse bias.*

An important consideration when comparing TA spectra results with ssPL, JV and TDCF is the relatively higher excitation density in TA measurements. It has been shown that singlet-singlet annihilation (SSA) already sets in at an exciton density of $1\times10^{16}$ cm$^{-3}$, which is a factor of ca. 13 lower than that when exciting our samples at a fluence of 2 μJ/cm$^2$.[51] This is also why the free charge generation efficiency in TAS is lower than that in TDCF; SSA competes with exciton dissociation in TAS but not in TDCF. We, therefore, complemented our studies with bias-dependent transient PL (trPL) measurements on both Y5 blends, with the results shown in **Figure 5c** and **SI Figure S14a** for L:Y5 and H:Y5 devices, respectively. For both blends, the trPL transients display two distinct time regimes: an early regime up to ca. 5 ns, in which the increasing reverse bias progressively speeds up the trPL signal decay, followed by a regime at longer timescales where the magnitude of signal decay shows a strong bias-dependence while the decay rate is weakly affected. Both blends exhibit similar decay properties, but the bias-dependence on the trPL shape and magnitude is strongly pronounced for the L:Y5 blend. By comparing the trPL kinetics of the L:Y5 blend with that of PS:Y5, we see that the second trPL decay regime of the blend corresponds exactly to the longer time decay of neat Y5 excitons, which have an average lifetime of 2 ns (see **SI Figure S14b**). This rules out that exciton reformation from long-lived CT states contributes significantly to the trPL signal in the considered ns time range.

### e) Analytical Models, QM Calculations and Simulations

From the above data, we arrive at the following conclusions:

i. There is no evidence for the build-up of an appreciable CT state population in the process of free charge generation, wherein geminate CT recombination would compete with CT separation. We conclude this from (a) the anticorrelation between the NFA LE ssPL in the blend and the free charge generation efficiency from TDCF, (b) the fact that the growth of the PM6 GSB goes nearly in parallel with the appearance of an EA signal in TAS (**see SI Figure S11c**), and (c) the distinct similarity in longer time trPL decay characteristics of the NFA when blended with a donor or dispersed by PS matrix.

ii. There is no evidence for LE-CT hybridization, which would provide the LE state with a stronger CT character affecting its emission properties.[54]–[56] This conclusion is reached based on (a) the excellent agreement of the PLQY of the blend, extrapolated to zero free charge generation, with the PLQY of the respective NFA in an inert PS matrix (see **SI Figure S15**), but also on (b) the lack of early charge generation in the poorly-performing system in TA measurements.

For all systems that obey the above two conditions, it must be possible to predict the absolute PLQY of the blend ($PLQY_{D:A,pred}$) from the field-dependence of the free charge generation efficiency and the probability that free charge recombination reforms LE excitons, according to:

$$PLQY_{D:A,pred}(E) = PLQY_{PS:A}\big(1 - IGE(E)\big) + ELQY_{D:A}J_{NGR,norm}(E) \qquad (2)$$

The elements of this model are illustrated in a simplified three-state diagram in **Figure 6a**. The first term denotes photon emission from local excitons on the NFA, which decay back to the ground state in competition with their separation into free charge. This is illustrated by emission pathway (i) in **Figure 6a**. From optical simulations and our comparison of the EGE with the saturation photocurrent of H:Y6, we conclude that 80 % of the incident photons are absorbed in our PM6:Y6 blends. We can, therefore, approximate the probability that an absorbed photon generates a free charge carrier by the field-dependent internal generation efficiency $IGE(E) = EGE(E)/0.8$. Since our experiments did not reveal any evidence for hybridization of local excitons and CT states, we then set the radiative efficiency of undissociated NFA excitons in the D:A blends equal to $PLQY_{PS:A}$, i.e. the PL quantum efficiency in the absence of the donor, which was earlier found to be field-independent. Accordingly, the fraction of initially absorbed photons reemitted from undissociated LE states against the competition with free charge generation is given as $PLQY_{PS:A} \cdot (1 - IGE(E))$. The second term describes NFA photon emission in the blends by the reformation of LE states from free charge recombination.[11], [15] The probability that an absorbed photon results in re-emission from a reformed NFA LE state is given by $ELQY_{D:A} \cdot J_{NGR,norm}(E)$. Here, $ELQY_{D:A}$ is the EL quantum efficiency of the blend and $J_{NGR,norm}$ is the non-geminate recombination current density normalized to the saturation current density $J_{gen}$. In our PM6:Yx blends, we set $J_{gen}$ equal to 27 mA/cm$^2$ as outlined earlier, which represents the case that every photogenerated exciton creates a free charge carrier.

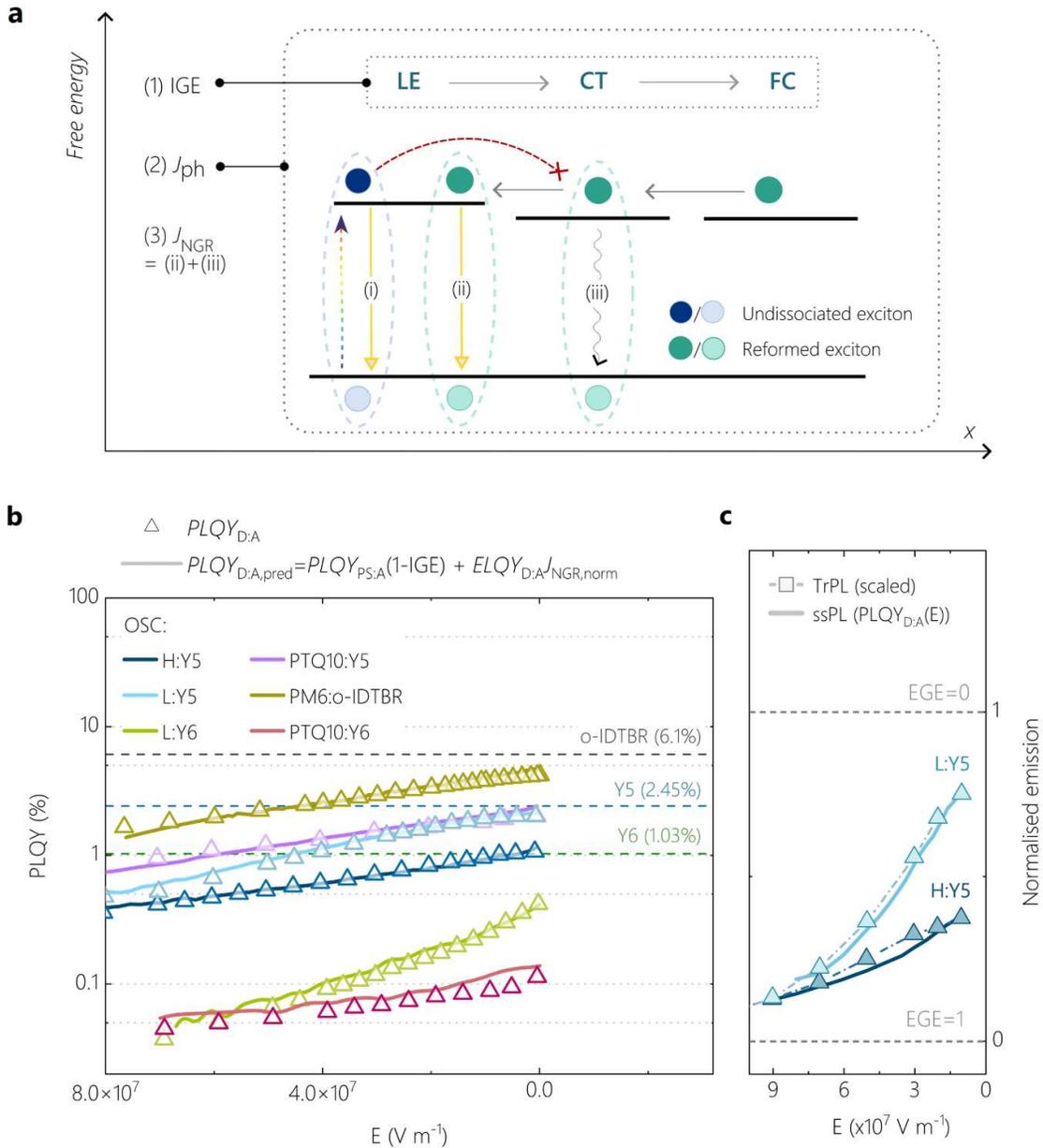

**Figure 6. Accurate reconstruction of experimental PLQY using a singlet-decay-limited generation model.** *(a) An illustration of the physical significance of $J_{ph}$, EGE and non-geminate bimolecular recombination current $J_{NGR}$. The EGE is the product of forward processes LE dissociation and CT separation into free charge (FC) states. Various recombination pathways that can occur also are illustrated in the Jablonski scheme. The blue circles represent undissociated LE excitons on the NFA that decay in competition with free charge generation (emission pathway (i)). Green circles represent exciton states formed by free charge carriers that undergo NGR: either via CT states (emission pathway (iii)) or via further LE state reformation (emission pathway (ii)). (b) For all test OSCs, a comparison of the field-dependence of the predicted $PLQY_{D:A,pred}$ (solid lines) reconstructed using equation (2), and the experimental $PLQY_{D:A}(E)$ (symbols) obtained from field-dependent ssPL and equation (1). (c) Agreement between the steady-state normalised $PLQY_{D:A}(E)$ and the field-dependence of trPL quenching, which is reconstructed from field-dependent exciton splitting efficiencies for L:Y5 and H:Y5 (see supplementary note 2).*

**Table 1.** *Comparison of the measured $PLQY_{D:A}$ and predicted $PLQY_{D:A,pred}$ at the built-in voltage ($V_{Bi}$) or zero field conditions. The latter is predicted from our model of singlet-decay limited free charge generation as outlined in the text. Also tabulated are the ELQY values used in equation (2). The lowermost row, obtained from the ratio of*

*ELQY of the blend to PLQY of PS:NFA , denotes the probability that non-geminate free charge recombination (NGR) proceeds via the reformation and decay of the NFA singlet.*

| in % | H:Y6 | L:Y6 | H:Y5 | L:Y5 | PM6:o-IDTBR | PTQ10:Y6 | PTQ10:Y5 |
|---|---|---|---|---|---|---|---|
| $PLQY_{D:A}$ | $5.2 \times 10^{-2}$ | 0.42 | 1.07 | 2.03 | 4.20 | 0.11 | 2.01 |
| $PLQY_{D:A,pred}(V_{Bl})$ | $5.9 \times 10^{-2}$ | 0.43 | 1.20 | 2.14 | 4.71 | 0.14 | 2.41 |
| $ELQY$ | $4.0 \times 10^{-3}$ | $3.5 \times 10^{-2}$ | 0.22 | 0.13 | 0.20 | $9.9 \times 10^{-3}$ | 0.38 |
| *Likelihood of NGR proceeding via NFA LE states* | 0.4 | 3.4 | 8.9 | 5.3 | 3.2 | 1 | 15.5 |

**Figure S15** in **SI** graphically shows the experimental values of IGE and $J_{NGR,norm}$ as a function of the electric field $E$ for the present four systems, used for the prediction of the blend's $PLQY_{D:A,pred}$ in equation (2). All other parameters are given in Table 1. As shown in **Figure 6b**, the predicted $PLQY_{D:A,pred}$ (denoted by hollow triangles) shows excellent agreement to the experimental $PLQY_{D:A}$ data (solid lines) over the entire field range and for all of our model systems. To further check for the generality of our conclusions, we extended our approach to other low-offset blend systems. We have recently shown that the blend of another high molecular weight batch of PM6 with the NFA o-IDTBR exhibits field-dependent exciton dissociation.[11] According to recent CV measurements and DFT calculations, this NFA exhibits an IE that is smaller than that of Y5 by more than 0.1 eV.[27] Indeed, our JV and TDCF measurements of this blend reveal an even more pronounced field-dependence of photogenerated charge and photocurrent than the L:Y5 system. We also blended the Y6 and Y5 acceptors with another donor polymer PTQ10, which, according to CV, PES and PESA, has an IE ca. 0.12 eV on average higher than $H_{PM6}$, rendering it a proper comparison with our $L_{PM6}$-based blends.[27] The absorption, photovoltaic performance, TDCF and TAS results for PTQ10:Y5 and PTQ10:Y6 are shown in detail in **SI Figures S16-S18**. Clearly, the PTQ10:Y5 blend exhibits field-assisted free charge generation characteristics similar to the L:Y5 blend, despite the very different chemical structure and molecular weight of PTQ10 and $L_{PM6}$. Notably, for all these various blends exhibiting different $\Delta$IE, we can closely describe their emission properties with our model assuming a free charge generation process that is primarily limited by singlet exciton decay, see **Figure 6b.** To substantiate this, **Table 1** shows the comparison of measured and predicted PLQY of the blend at $V_{Bl}$ conditions for all our tested OSCs. Interestingly, the contribution to the total emission from LE states which are reformed via NGR is very minor (see **SI Figure S19**). From **Table 1**, the ratio of $ELQY_{D:A}$ to $PLQY_{PS:A}$ describes the likelihood that non-geminate free charge recombination occurs through emissive LE states, and this is found to be smaller than 10% (in most cases, less than 5%). In fact, the main difference between LE dissociation into free charge and free charge recombination via LE states is that the latter process produces a significant density of triplet CT states, which can decay further to local triplet excitons through back electron transfer.[57] Recent TA work suggested that free charge recombination in PM6:Y6 proceeds almost entirely through this pathway[58], which our results indeed agree with.

The critical role of LE splitting efficiency in blend emission is also evidenced by trPL. **SI Figure S20** shows the $S_1$ splitting efficiency as a function of applied bias for the Y5-based blends, obtained by integrating the trPL kinetics of each blend. If the reduction of $PLQY_{D:A}$ with increasing reverse bias originates entirely from field-induced quenching of the singlet exciton emission, the trPL quenching efficiency as function of electric field should align perfectly with the $PLQY_{D:A}$ data (see Supplementary Note 2 for description). This is indeed the case and is shown in **Figure 6c** for the two PM6:Y5-based blends. The trPL quenching of the Y5-based blends also show convergence at high external bias, which additionally confirms our conclusion that the loss channel *via* radiative decay of the LE state can be substantially circumvented at high field conditions, despite smaller ΔIE in the inefficient blends.

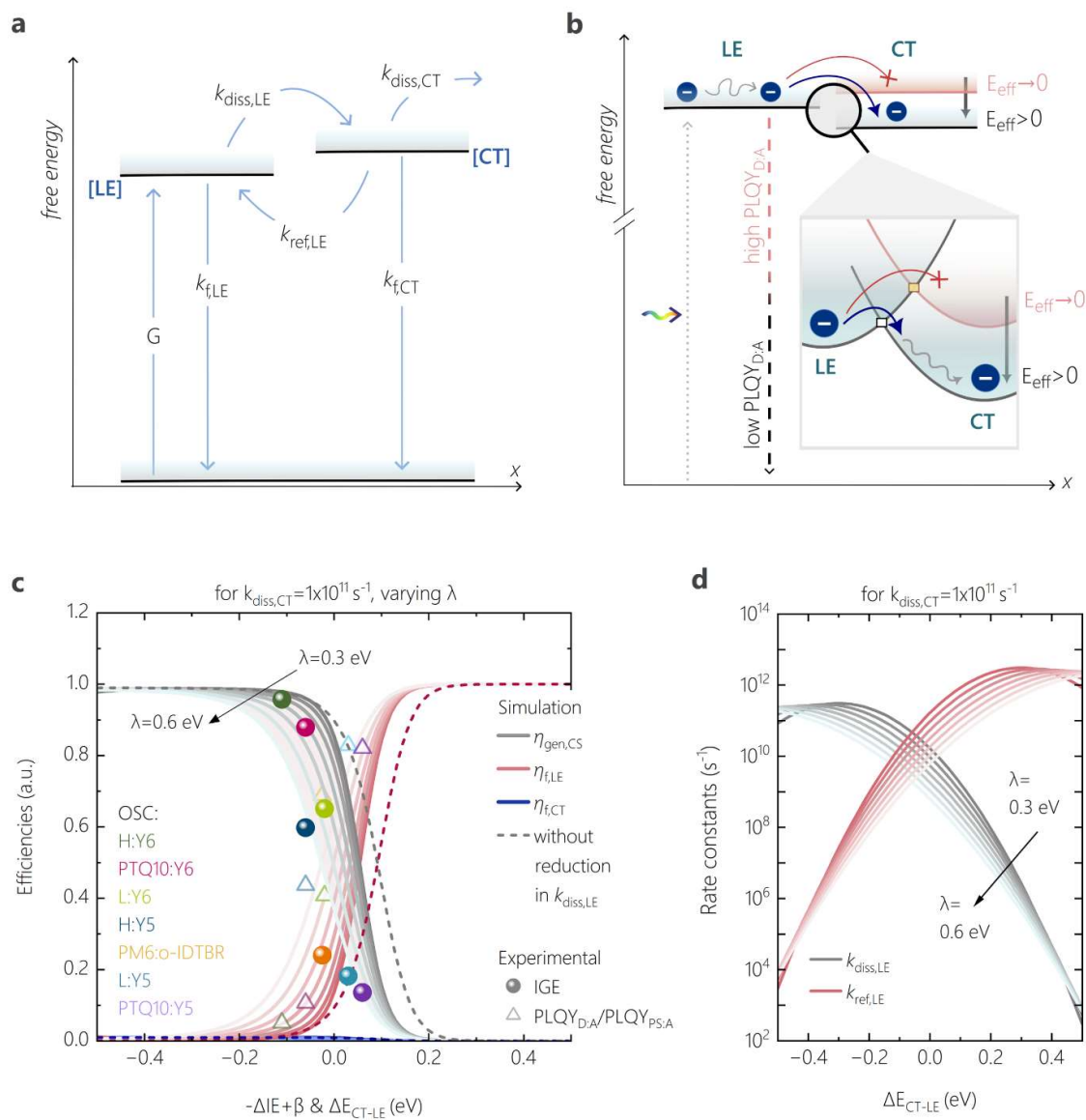

**Figure 7.** *Marcus theory used to explain singlet exciton decay as the competing pathway to free charge generation. (a) Rate model showing the various transitions possible for excitons prior to free charge formation. (b) A representation of the LE dissociation based on Marcus theory. (not to scale) The dark blue circles depict bound charge carriers in their excitonic states, and energetic states under no field (applied field) are described with red (dark grey) lines, curves and text, respectively. The barrier for LE to CT formation is explained by the Marcus type energy picture at the interface, highlighted in the box. The squares on the potential curves denote the cross-over point between LE and CT energetic potentials, and signify the activation barrier that bound charge carriers in vibrationally relaxed LE states must overcome to form the CT state – a barrier that is lowered under the influence of an effective field. (c) The dependence of charge generation efficiency, losses via the decay of the local NFA exciton and the CT state ($\eta_{gen,CS}$, $\eta_{f,LE}$, $\eta_{f,CT}$, respectively) on $\Delta E_{CT-S_1}$, simulated at zero-field from the steady-state rate model assuming not all photogenerated excitons are able to undergo charge transfer. The simulated data is shown for varying reorganisation energies λ. Overlaid are the experimental generation efficiencies IGE and emission efficiencies of various tested blends (ratio of blend to NFA PLQY) measured at open-*

To quantitatively address the role of the IE offset on the efficiency of free charge formation, we solved the standard rate model for the steady state population of LE and CT states[12], [59], as illustrated in **Figure 7a:**

$$\frac{d[LE]}{dt} = G + k_{ref,LE}[CT] - k_{diss,LE}[LE] - k_{f,LE}[LE] = 0 \tag{3a}$$

$$\frac{d[CT]}{dt} = k_{diss,LE}[LE] - k_{diss,CT}[CT] - k_{f,CT}[CT] - k_{ref,LE}[CT] = 0 \tag{3b}$$

Here, $[LE]$ and $[CT]$ are the densities of LE and CT, respectively. $G$ is the generation rate of LE states, $k_{f,LE}$ and $k_{f,CT}$ are the decay rates of LE and CT excitons, respectively, $k_{diss,LE}$ and $k_{diss,CT}$ are the dissociation rates of LE and CT excitons, respectively, and $k_{ref,LE}$ is the reformation rate of LE states from CT states.

Since we are only concerned with the generation pathway, we exclude the reformation of CT states from free charge carriers through NGR. We then used Marcus theory to calculate the rates $k_{diss,S_1}$ and $k_{ref,LE}$ as illustrated in **Figure 7a:**

$$k_{diss,LE} = 0.1 \frac{|H_{DA}|^2}{\hbar} \sqrt{\frac{\pi}{\lambda k_B T}} exp\left(-\frac{(\Delta E_{CT-LE} + \lambda)^2}{4\lambda k_B T}\right) \tag{4a}$$

$$k_{ref,LE} = \frac{|H_{DA}|^2}{\hbar} \sqrt{\frac{\pi}{\lambda k_B T}} \exp\left(-\frac{(-\Delta E_{CT-LE} + \lambda)^2}{4\lambda k_B T}\right) \tag{4b}$$

Here, $|H_{DA}|$ is the electronic coupling of the donor and acceptor for hole transfer, $\lambda$ the corresponding reorganization energy, $k_B$ the Boltzmann constant, and $T$ the temperature. The energetics at the heterojunction are expressed by the energetic offset between the CT and LE states ($\Delta E_{CT-LE}$) which is related to the ΔIE in a first order approximation, as mentioned earlier. Following common practice, we assumed that the effective density of states of local excitons in our blends is ten times that of CT states[14], meaning that not every excited NFA has a neighbouring donor molecule to undergo charge transfer. This rationalizes the pre-factor 0.1 in equation 4a. In contrast, every CT state has at least one neutral neighbouring NFA molecule available for LE reformation.

To ascertain the values of $|H_{DA}|$ and the inner reorganization energy for charge transfer, quantum mechanical (QM) calculations were performed on model D:A interface systems. The computational modelling therein was also used to atomically resolve and compare the nature of the neutral LE and CT states of the interfacial systems and to determine the excited state energies Here, the interface systems consist of either a single Yx molecule or an optimized Yx dimer stacked onto a PM6 oligomer (of two repeat units) with a stacking distance of 4.5 Å. Further details can be in found in **SI Supplementary Note 3.** Figure 8 shows the natural transition orbitals (NTOs) of the electrons and holes of the excited states along with the excitation energies ($E_{LE}/E_{CT}$), oscillator strength (*f*) and weight of CT character (CT%) of the Yx-dimer/PM6 interfacial system (denoted here as an interfacial trimer or 2Yx+PM6). **SI Figure S21 and Table S2-S3** provides the data for the corresponding interfacial systems with only one Yx molecule (interfacial dimer or Yx+PM6). The LE states of the interfacial trimer in Y5 and Y6-based systems in Figure 8a and 8b have similar excitation energy, $E_{LE}$=1.790 eV vs. $E_{LE}$=1.770 eV, wherein the corresponding NTO hole and electron MOs are also very similar to each other. The predicted LE state energies are still much higher (>0.30 eV) than what we have found in the experiment. In our previous work, we nicely predicted the absorption spectra of Y6 aggregate, where the energy went from 2.06 eV for the Y6 monomer down to 1.48 eV for the Y6 aggregate.[60]. As we concluded before, this is mainly due to the delocalization effect of excitonic states in the Y6 aggregate. We can observe this effect in the NTOs LE state figures in Figure 8a and 8b, where the MOs are delocalized to the next Yx molecule in the 2Yx+PM6 interfacial systems, compared to the Yx+PM6 systems in Figure S21. As expected, and nicely shown by the comparison of the corresponding NTOs in Figure 8 and Figure S21, delocalization effects are much weaker for the CT states. The only difference between the two systems is the "F" atoms in Y6 molecules being replaced by "H" atoms to become Y5 molecules. The electron affinity difference of the two acceptor molecules also changes the electron population of the PM6, especially in the region near the extra fluorination site.

In detail, we find that the LE transition energy is about 25-30 meV larger for the Y5-based interfacial dimer and trimer systems compared to the Y6-containing systems, see **SI Table S2**. This is significantly smaller than the difference of about 100 meV in the peak position in absorption of the NFAs (see Figure 1d and SI Figure S2b). We attribute this to the stronger tendency of Y6 to aggregate. The CT energy of the Y5-based aggregates is predicted to be more than 100 meV higher than the Y6 counterparts, see **SI Table S2**, which is consistent with the approximately 100 meV higher HOMO and LUMO energies and corresponds nicely to the almost 100 mV higher $V_{OC}$ of the Y5-based blends (see Figure 2c). Regarding $\Delta E_{CT-LE}$, the LE state lies above the CT state in all calculated systems, but the energy difference decreases significantly with aggregate size. The reason for this is the noted decrease of the LE energy with an increasing number of NFAs in the stack which is in contrast to the energy of the CT state which changes only little with aggregate size. We, therefore, expect $\Delta E_{CT-L}$ to eventually tend closer to zero or to even adopt positive values in a real D:A aggregate, especially in the low offset blends.

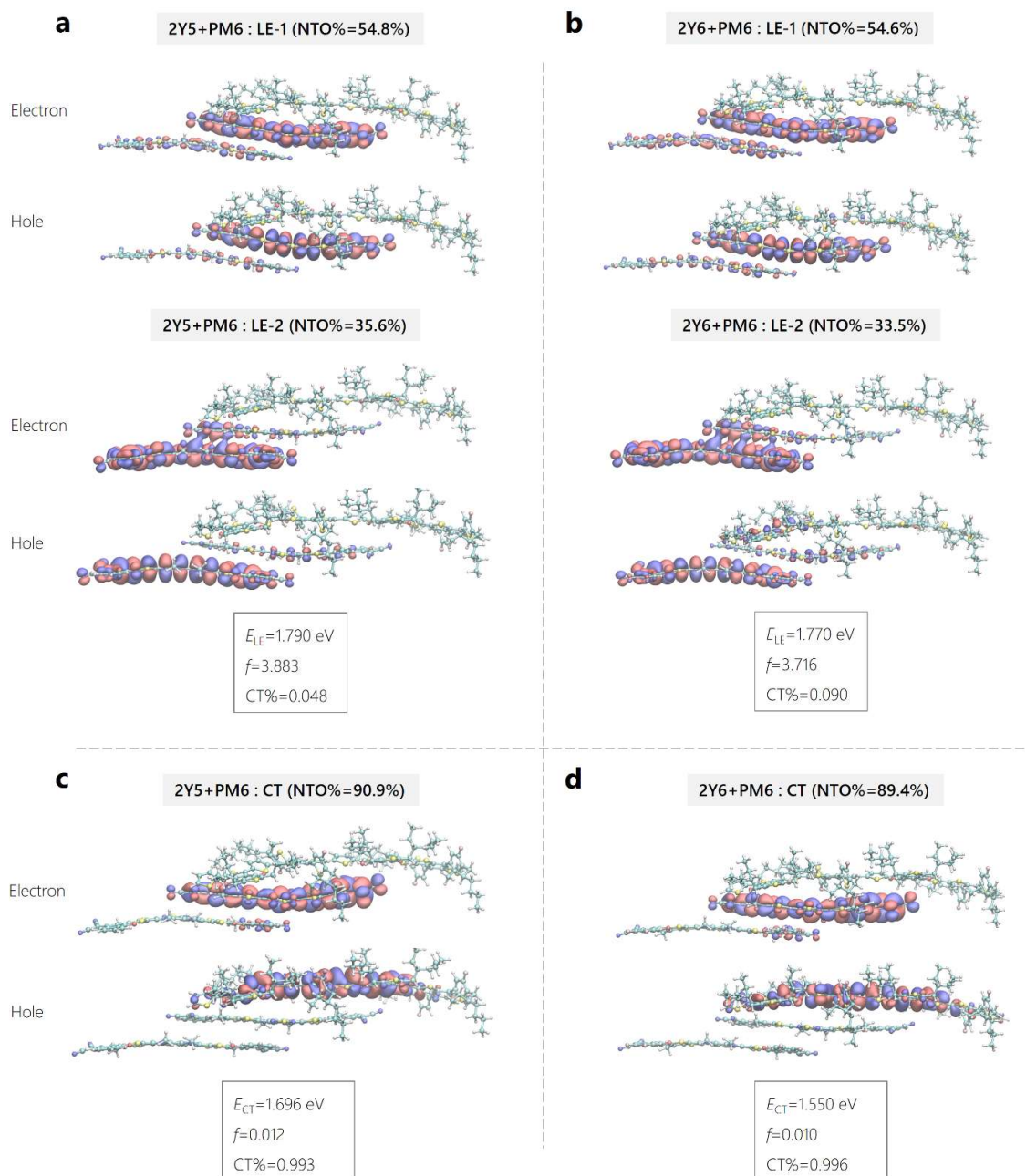

**a**  2Y5+PM6 : LE-1 (NTO%=54.8%)

Electron

Hole

**b**  2Y6+PM6 : LE-1 (NTO%=54.6%)

Electron

Hole

2Y5+PM6 : LE-2 (NTO%=35.6%)

Electron

Hole

2Y6+PM6 : LE-2 (NTO%=33.5%)

Electron

Hole

$E_{LE}$=1.790 eV
$f$=3.883
CT%=0.048

$E_{LE}$=1.770 eV
$f$=3.716
CT%=0.090

**c**  2Y5+PM6 : CT (NTO%=90.9%)

Electron

Hole

**d**  2Y6+PM6 : CT (NTO%=89.4%)

Electron

Hole

$E_{CT}$=1.696 eV
$f$=0.012
CT%=0.993

$E_{CT}$=1.550 eV
$f$=0.010
CT%=0.996

*Figure 8. Modelling of excited state energies for $H_{PM6}$:Yx geometries. (a-b) Natural transition orbitals or NTOs (hole and electron) of the singlet local excited (LE) state in the 2Y5+PM6 and 2Y6+PM6 system. (c-d) NTOs of the charge transfer (CT) states for the same Y5- and Y6-based interfacial trimer systems. The two pairs NTOs of the LE state and one pair NTOs of the CT state are plotted, along with the corresponding weights (NTO%). The excitation energies ($E_{LE}/E_{CT}$), oscillator strength (f) and weight of CT character are provided as well.*

**Table 2** summarizes all parameters which were used as constants in the steady-state rate model. The reorganisation energy $\lambda$ consists of the inner and outer reorganization energy, $\lambda_i$ and $\lambda_a$, respectively, where the former is mainly related to the change of the molecular geometry, while the latter originates from the change in the polarization of the environment during charge transfer. Values for $\lambda_i$ were obtained from computational calculations (see SI Table S3 and SI Supplementary Note 3 for more details) and are similar for both Y5- and Y6 aggregates (ca. 325-330 meV). The value of $\lambda_a$ ($\cong$ 250 meV) was taken from literature.[61] This yields $\lambda = \lambda_i + \lambda_a$ of ca. 580 meV. This value is slightly higher than the value predicted for a PBDB-T/ITIC dimer in literature.[62] Note that contrastingly smaller values for the inner (161 meV) and outer (150 meV) reorganization energy were calculated for a J61:m-ITIC dimer.[63] We, therefore, varied $\lambda$ between 300 and 600 meV in steps of 50 meV. $|H_{DA}|$ was set equal to 10 meV based on the average of the results from DFT. The emissive decay rates $k_{f,LE}$ and $k_{f,CT}$ were both set equal to 1×10$^9$ s$^{-1}$, a typical value for our and other state of the art NFAs[14] and for the CT decay time[64], respectively. Given the lack of recombination through the CT state during free charge formation, rapid CT dissociation was assumed which is driven, for example, by band bending at the DA heterojunction[12]. We note here that $k_{diss,CT}$ stands for the split-up of the interfacial CT state and that its final separation into free charge carriers may take longer times.

*Table 2. Fixed parameters used in the steady-state model for LE and CT populations.*

| Parameter | | Value |
|---|---|---|
| Decay rate of NFA LE states | $k_{f,LE}$ [$s^{-1}$] | $1 \times 10^9$ |
| Decay rate of CT states | $k_{f,CT}$ [$s^{-1}$] | $1 \times 10^9$ |
| Dissociation rate of CT states | $k_{diss,CT}$ [$s^{-1}$] | $1 \times 10^{11}$ |
| Generation rate of LE states | $G$ [$m^{-3}s^{-1}$] | $1 \times 10^{28}$ |
| Total reorganization energy | $\lambda$ [eV] | 0.3 to 0.6 |
| D:A electronic coupling for hole transfer | $|H_{DA}|$ [eV] | 0.01 |

The populations $[LE]$ and $[CT]$, obtained from equation 3a and 3b and the parameters in Table 2 in steady-state condition, were used to calculate three efficiencies according to equation 5a-5c:

$$\eta_{gen,CS} = \frac{k_{diss,CT} [CT]}{G} \tag{5a}$$

$$\eta_{f,LE} = \frac{k_{f,LE} [LE]}{G} \tag{5b}$$

$$\eta_{f,CT} = \frac{k_{f,CT} [CT]}{G} \tag{5c}$$

Here, $\eta_{gen,CS}$ stands for the free charge generation efficiency while $\eta_{f,LE}$ and $\eta_{f,CT}$ describe the losses via the decay of the NFA LE state and the CT state, respectively. These three efficiencies are plotted against $\Delta E_{CT-S1}$ in **Figure 7c**, for a $k_{diss,CT}$ of 1x10$^{11}$ s$^{-1}$. We observe a sharp decrease in $\eta_{gen,CS}$ when the CT state moves above the LE state ($\Delta E_{CT-LE} > 0$), as also shown by other works. [12], [59] We note two other findings that are important here. First, the LE decay – as displayed by the red curves in Figure 7c – is shown to be the main competing channel to free charge generation. This is a direct consequence of the choice of a high CT dissociation rate, as shown in **SI Figure S22**. Second, there exists a critical $\Delta E_{CT-LE}$ above which $\eta_{gen,CS}$ is a strong function of $\lambda$. Interestingly, the overall dependence of $\eta_{gen,CS}$ on $\Delta E_{CT-LE}$ and $\lambda$ is similar to that of $k_{diss,LE}$, which is shown with grey lines in Figure 7d. To understand this, we analysed the analytical solution of equations 3a and 3b in the Supporting Information for different cases (please refer to SI Appendix Note for further discussion and explanation). It is exactly for the case of fast CT dissociation (case 1 in the SI Appendix Note) that $\eta_{gen,CS}$ is determined by $k_{diss,LE}$, which is a strong function of $\lambda$ through equation 4a. On the other hand, if CT dissociation is slow, a kinetic equilibrium between the CT and the LE population is established, where $[CT]/[LE]$ is governed by $k_{diss,LE}/k_{ref,LE}$ (case 2 in the SI Appendix). Since $k_{diss,LE}/k_{ref,LE}$ is independent of $\lambda$ (see **SI Figure S23**), the influence of $\lambda$ on $\eta_{gen,CS}$ becomes reduced, as shown in **SI Figure S22a**. Finally, if the exciton dissociation rate increases and dominates the LE decay rate (i.e. $k_{diss,LE}/k_{ref,LE} \cong k_{diss,LE}$), the CT dissociation probability determines $\eta_{gen,CS}$, albeit with a smaller influence of $\lambda$ (case 3 in SI Appendix and **SI Figure S24**).

To compare with experimental values, we considered IGE at $V_{OC}$ (as measured by mTDCF) as the generation efficiency and the measured $PLQY_{D:A}(V_{OC})/PLQY_{PS:A}$ as the radiative decay efficiency of LE states. We used a constant scaling factor $\beta$ to relate the experimental $\Delta$IE from SEC to $\Delta E_{CT-LE}$ in the Marcus rate calculations, using $\Delta E_{CT-LE} = -\Delta\text{IE} + \beta$. Such scaling factors were used in the past to account for differences between the LE and CT binding energy but also possible band-bending which lifts $E_{CT}$ relative to $E_{LE}$.[9] In our work, the $\Delta$IE was scaled to reproduce an $\Delta E_{CT-}$ of -0.11 eV for H:Y6, which was determined previously from temperature-dependent ELQY[30] (see SI Supplementary Note 4 for further details on this). **Figure 7c** shows an excellent agreement when overlaying the experimental values with the simulation result. In particular, the predicted drop in $\eta_{gen,CS}$ and the rapid increase in $\eta_{rec,\text{LE}}$ upon the small variation in $\Delta E_{CT-L}$ (by 200 meV) is especially consistent with the experimental anticorrelation between the blends' emission and free charge generation data. Our combined experimental-simulation work shows that a $\Delta E_{CT-LE}$ of ca. -0.1 V, corresponding to $\Delta$IE of ca. 0.35 eV, is needed for efficient free charge generation against the competition with LE recombination. This is a direct consequence of the dominant role of the LE dissociation rate and it is related to $\Delta E_{CT-L}$ and also to $\lambda$.

Our model also explains the pronounced effect of the electric field on the exciton spitting efficiency in our low offset systems. If the CT state can be easily polarized, it stands that an external field stabilizes the CT state thereby lowering its energy relative to that of the LE state[65], as illustrated in the inset of **Figure 7b**. This diminishes the uphill barrier to CT formation and promotes free charge generation by facilitating LE dissociation. Consequentially, the radiative decay via NFA LE states depends on the ability of such photogenerated LE states to contribute to free charge generation, see **Figure 7b**. This charge transfer process enabled by the internal field directly explains the PLQY trends predicted for different external bias, and is likely the key factor impacting photocurrent in low-offset NFA based OSCs.

Finally, we would like to address the role of the morphology. Numerous studies have shown that charge separation benefits from a well phase-separated structure with rather pure and possibly well-crystallized domains of the donor and acceptor components.[66]–[68] It has been previously suggested that such a morphology reduces the CT binding energy either through delocalization of the interfacial CT state, or by providing a larger driving force to counteract the mutual Coulombic attraction of the electron-hole pair towards the charge separated state.[12], [39], [65], [69], [70] However, charge separation does not appear to be a major obstacle in the systems investigated here, as evidenced from the anticorrelation of TDCF and PL data, TA measurements and analytical simulations. On the other hand, LE dissociation takes place on a smaller length scale. Thus, although we acknowledge that the intermolecular order and orientation at the D:A heterojunction can affect critical parameters for LE-CT interactions,[29], [71], [72] $\Delta IE$ ($\Delta E_{CT-LE}$) seems to be the most important property for the free charge generation process. A systematic dependence of the device performance on $\Delta IE$ ($\Delta E_{CT}$) was indeed reported for a wide range of NFA-based systems, in both planar heterojunctions and bulk heterojunctions, with very different morphologies.

## Conclusion

To conclude, our simulations and experiments consistently reveal a large effect of the IE offset on the free charge generation efficiency in low-offset OSCs, which is hindered by CT formation from LE states. Our key finding is that as the energy offset reduces, the decay of NFA singlet excitons becomes the primary and direct competition to free charge generation. We also learn that the IE offset of PM6:Y6 is close to the limit of the region where the exciton dissociation efficiency is close to one. In agreement to this, all state-of-the-art high efficiency binary blend devices have been based on the combination of Y6 (or derivatives of this NFA with slightly different chemical structure but similar IE) with PM6, D18 or polymers with related chemical structure and energetics.[1], [73], [74] Therefore, our conclusion also encourages further material design along similar lines and suggests that any attempts to further reduce the IE offset in the blend will lead to an unavoidable loss in free charge generation efficiency. However, if the IE offset is in fact reduced, then Marcus theory provides guidelines so that the solar cell could still balance high photovoltage with high photocurrent. For example, a reduction of the reorganization energy for charge transfer will enable efficient free charge generation even for $\Delta E_{CT-S1}$ close to zero (in Figure 7c, corresponding to ΔIE of ca. 240 meV). For example, Zhong *et al.* reported sub-ps charge transfer in ITIC based donor:NFA blends which was related to the small reorganization energy for charge transfer.[63] In addition, a smaller reorganization energy for hole back transfer will reduce the non-radiative voltage loss due to electron-vibration coupling.[75] Recent works have demonstrated the power of these strategies to obtain highly efficient single-junction OSCs albeit with smaller ΔIEs than in PM6:Y6.[76], [77]

## Conflicts of interest:

There are no conflicts of interest to declare.

## Acknowledgements:

We acknowledge funding from the Deutsche Forschungsgemeinschaft (DFG, German Research Foundation) through the project Fabulous (Project Number 450968074) and project Extraordinaire (Project Number 460766640). D.M., J.R. and S.L. acknowledge financial support from the DFG via GRK2948 and the Center for Integrated Quantum Science and Technology (IQST), University of Stuttgart. K.V. acknowledges financial support from the European Research Council (ERC, grant agreement 864625). S.S., W.M. and K.V thank FWO Vlaanderen for financial support (1S99620N).

## On the critical competition between singlet exciton decay and free charge generation in non-fullerene based organic solar cells with low energetic offsets

Manasi Pranav[1], Atul Shukla[1], David Moser[2], Julia Rumeney[2], Wenlan Liu[3], Rong Wang[4], Bowen Sun[1], Sander Smeets[5,6], Nurlan Tokmoldin[1,7], Frank Jaiser[1], Thomas Hultzsch[1], Safa Shoaee[1,7], Wouter Maes[5,6], Larry Lüer[4], Christoph Brabec[4,8], Koen Vandewal[5,6], Denis Andrienko[3], Sabine Ludwigs[2], Dieter Neher[1*]

### Affiliations:

[1] Institute of Physics and Astronomy, University of Potsdam, Karl-Liebknecht Straße 24/25, 14476 Potsdam, Germany

[2] IPOC – Functional Polymers, Institute of Polymer Chemistry, University of Stuttgart, Pfaffenwaldring 55, 70569 Stuttgart, Germany

[3] Max Planck Institute for Polymer Research, Ackermannweg 10, 55128 Mainz, Germany

[4] Institute of Materials for Electronics and Energy Technology (i-MEET), Friedrich-Alexander-Universität Erlangen-Nürnberg, Martensstrasse 7, Erlangen 91058, Germany

[5] UHasselt—Hasselt University, Institute for Materials Research, (IMO-IMOMEC), Agoralaan 1, 3590 Diepenbeek, Belgium

[6] IMOMEC Division, IMEC, Wetenschapspark 1, 3590 Diepenbeek, Belgium

[7] Heterostructure Semiconductor Physics, Paul-Drude-Institut für Festkörperelektronik, Leibniz-Institut im Forschungsverbund Berlin e.V, Hausvogteiplatz 5-7, 10117 Berlin, Germany

[8] Helmholtz-Institut Erlangen-Nürnberg for Renewable Energies (HIERN), Forschungszentrum Jülich, Immerwahrstraße 2, 91058 Erlangen, Germany

*Corresponding author: neher@uni-potsdam.de

### Experimental methods

#### Fabrication:

For Y5-based blends, the commercial polymer PM6 with high molecular weight and the small non-fullerene acceptor Y5 were both purchased from 1-Materials Inc. For Y6-based blends, the commercial PM6 polymer and small non-fullerene acceptor Y6 were purchased from Brilliant Matters Inc. The low molecular weight PM6 synthesis was conducted on a custom-made continuous flow system. Exact synthesis conditions have been reported previously.[1] Flow rates were set to achieve a residence time of 1.6 min to obtain the targeted number-average molar mass ($M_n$) of 3.5 kg/mol.

Although optimised PM6:Y6 and PM6:Y5 (with high $M_n$) devices are typically fabricated with a PDINO electron transport layer,[2] we opted to use PDINN for our studies. While this leads to lower PCEs, PDINN possesses higher conductivity due to doping[3] which enables shorter dielectric relaxation times and a faster response of the device in transient optoelectronic measurements, such as TDCF. Devices were fabricated in a conventional geometry with a structure ITO/PEDOT:PSS/D:A/PDINN/Ag. Glass substrates with pre-patterned ITO (Lumtec) were cleaned in an ultrasonic bath with acetone, Hellmanex III, deionized water and isopropanol for 10 minutes each, followed by microwave oxygen plasma treatment (4 min at 200 W). Subsequently, an aqueous solution of PEDOT:PSS (Heraeus Clevios™ PEDOT:PSS) was filtered through a 0.2 µm PTFE filter and spin coated onto ITO at 5000 rpm under ambient conditions. The PEDOT:PSS coated substrates were thermally annealed at 150 °C for 15-20 min. Blends solutions were prepared to a total concentration of 14 mgmL$^{-1}$ (17 mgmL$^{-1}$ for low $M_n$-based blends) using a CHCl$_3$ solvent (purchased from Carl Roth), with a 1:1.2 weight ratio. The solution was stirred for 3 hours inside a nitrogen-filled glovebox. Polystyrene (Sigma Aldrich) blends with the NFAs were prepared similarly, with similar concentration and blend ratio as the D:A blends. The blend was spin coated onto the PEDOT:PSS coated substrates using the optimized spin speed for obtaining a photoactive layer of thickness of 100-110 nm. Y6-based blends were annealed at 100 °C for 10 min. Then, a 1 mgmL$^{-1}$ solution of PDINN (1-Material Inc) in methanol (Sigma Aldrich) was spin coated at 1500 rpm. Finally, 100 nm of silver as the top electrode was evaporated under a 10$^{-6}$-10$^{-7}$ mbar vacuum. The resulting active area was 1 mm$^2$ for TDCF and steady state biased-PL measurements, and 6 mm$^2$ for EQE$_{PV}$, JV, EL, ELQY, and TrPL measurements.

For PLQY measurements, glass substrates with full-area ITO were used to produce ca. 1 cm$^2$ device area, to ensure that the entire device stack is illuminated by the incident beam. For spectroelectrochemistry, the photoactive blends were spin coated on full-area ITO glass subtrates at reduced blend concentrations to obtain films of ca. 30 nm thickness. The preparation method for PM6:o-IDTBR OSCs has been previously reported.[4]

### Current density-voltage characteristics (JV)

JV curves were measured using a Keithley 2400 SourceMeter in a 2-wire source configuration. Simulated AM1.5G irradiation at 100 mWcm-2 was provided by a filtered Oriel Sol2A Class AA Xenon lamp and the intensity was monitored simultaneously with a Si photodiode. The sun simulator is calibrated with a KG5 filtered silicon solar cell (certified by Fraunhofer ISE).

### EQE$_{PV}$ and absorbance

The EQE$_{PV}$ was measured with broad white light from a 300 W Halogen lamp (Phillips) which was chopped at 80 Hz (Thorlabs MC2000), guided through a Tornerstone monochromator and coupled into an optical quartz fibre, calibrated with Newport Photodiode (818-UV). An SR 830 Lock-In Amplifier measures the response of the solar cell under different bias voltages applied by a Keithley 2400.

Absorbance was measured on films coated under the same fabrication conditions mentioned above, but on glass substrates with a Varian Cary 5000 spectrophotometer in transmission mode.

### Bias dependent photoluminescence (PL) and absolute PL/EL (PLQY/ELQY)

Bias-dependent PL measurements were performed using a 520 nm CW laser diode (Insaneware) for steady state illumination, and the intensity of the laser was adjusted to a 1 sun equivalent by illuminating a PM6:Y6 solar cell under short-circuit (provided by a Keithley 2400) and matching the current density reading to the $J_{SC}$ obtained in the sun simulator. The excitation beam was focused onto the sample using a stage of mirrors and lenses. Bias voltages ranging from open-circuit voltage to -8 V were applied to the sample using the same Keithley 2400. To ensure that only the active layer is illuminated and contributes to the emission response, we masked the measured pixels. The emission spectra were recorded with an Andor Solis SR393i-B spectrograph with a silicon DU420A-BR-DD detector and an Indium Gallium Arsenide DU491A-1.7 detector. A calibrated Oriel 63355 lamp was used to correct the spectral response. PL spectra were recorded with different gratings at central wavelengths of 800, 1100, and 1400 nm, and merged afterwards. For PLQY measurements, the same laser was used but the excitation beam was channelled through an optical fibre into an integrating sphere containing the sample. A second optical fibre was used from the output of the integrating sphere to the Andor Solis SR393i-B spectrograph. The spectral photon density was obtained from the corrected detector signal (spectral irradiance) by division through the photon energy ($h\nu$), and the photon numbers were calculated from the numerical integration, using a Matlab code.

For absolute EL measurements, a calibrated Si photodetector (Newport) connected to a Keithley 485 picoampere meter was used. The detector, with an active area of ~2 cm$^2$, was placed in front of the measured pixel with a distance <0.5 cm, and the total photon flux was evaluated considering the emission spectrum of the device and the external quantum efficiency of the detector. The injected current was monitored with a Keithley 2400.

## Time-delayed collection field (TDCF) and modified-TDCF (mTDCF)

In TDCF, the device was excited with a laser pulse from a diode pumped, Q-switched Nd:YAG laser (NT242, EKSPLA) with ~6 ns pulse duration at a typical repetition rate of 500 Hz. An Agilent 81150A pulse generator was used to apply a square voltage transient waveform constituting the pre-bias $V_{pre}$ and collection bias $V_{coll}$. To compensate for the internal latency of the pulse generator, the laser pulse was delayed and homogeneously scattered in an 85 m long silica fiber (LEONI) after triggering a photodiode. The device was illuminated while held at different pre-bias $V_{pre}$. After a pre-set delay time (calculated from the falling slope of the photodiode trigger), a high reverse bias $V_{coll}$ was applied to extract all the charges generated in the device. $V_{pre}$ and $V_{coll}$ were sent by the Agilent 81150A pulse generator through a home-built current amplifier, which was triggered by a fast photodiode (EOT, ET-2030TTL). The current flowing through the device was measured via a 10 Ω resistor in series with the sample and recorded with an oscilloscope (Agilent DSO9104H). To avoid non-geminate recombination of photogenerated free charge carriers prior to extraction, the intensity of light is kept very low and the delay time of collection is set to ~1ns.

For mTDCF, a square-type waveform was programmed and fed into the Agilent 81150A pulse generator, and the delay parameters for the $V_{pre}$ and $V_{coll}$ voltage steps were pre-defined prior to the measurement, such that the $V_{pre}$ was applied to the device for a duration equalling the RC time of the device plus the laser slew rate. As in regular TDCF, the laser beam incidence is delayed to compensate for the internal latency of the function generator.

## Spectroelectrochemistry

The spectroelectrochemical measurements were taken place with a three-electrode setup under argon atmosphere. The blends with ITO coated glass as base substrate were used as working electrode, a platinum wire as counter electrode and a silver chloride covered silver wire as pseudo reference electrode. All the potentials were referenced to the ferrocene/ferrocenium (Fc/Fc+) redox potential, added after the measurements. The electrochemical cell contain MeCN as solvent and TBAPF6 (0.1 molar) as conducting salt as standard electrolyte. The UV-vis measurements were performed in transmission mode through the working electrode. For each data point in the cyclic voltammogram, a spectrum was measured. The potentials were applied by an PGSTAT204 potentiostat, purchased by Metrohm. The UV vis spectra were measured by a spectrometer system purchased by Zeiss, containing a MCS621 vis II detector and a CLH600 F halogen lamp.

## Transient absorption (TA) spectroscopy

The femtosecond transient spectroscopy measurements were performed using a home-built setup. The output of an amplified Ti: Sapphire laser (Libra, Coherent, 800 nm, 1 KHz) was coupled to an Optical Parametric Amplifier (Opera Solo, Coherent) to generate the pump laser pulses (50 fs, 720 nm) which were chopped at $\omega/2$ (500 Hz). A portion of the fundamental laser beam was focused to an undoped YAG crystal to generate a broadband white light continuum. The probe pulses were spectrally dispersed using prism spectrometers and then collected using either a CMOS camera (visible spectral rang) or InGaAs photodiode arrays (IR spectral range). The probe continuum beam was split using a broadband beam splitter before the sample to monitor probe intensity on a shot-to-shot basis to minimize the impact of pulse-to-pulse fluctuation. The pump-probe polarization was set to magic angle (54.7°) to avoid orientational effects. The sequential probe shots corresponding to the pump on versus off were used to calculate the differential absorption signals. The pump-probe time delays up to ~7 ns are produced via a retroreflector connected to a computer controlled translational stage. In typical measurements, 6000 shots were averaged at each time delay and were repeated for at least 3 scans. The data saved as .dat files were processed via MATLAB for background and chirp correction. All the measurements were carried out under a nitrogen environment to prevent degradation from air contact.

## Time-resolved photoluminescence (trPL)

Photoluminescence decay and quenching in pristine and BHJ films were recorded using Picoquant Fluotime 300. A 402 nm pulsed laser was used to excite polymer donors and a 783 nm pulsed laser was used for Y5 and Y6 acceptors. The emission signals were detected at the wavelength with the maximum steady PL peak using a single-photon counting PMT. Data fitting was performed using the internal software EasyTau2. For the field dependent TrPL measurement, a holder was built to connect the device samples to a Keithley digital source meter to control the applied bias voltage from 0 V to −12 V.

## Computational modelling

We first set up a series of Y6 + Y6 + PM6 model trimer systems, where the Y6 dimer geometry was extracted from the optimized Y6 crystal structure [cite liu2023], and the PM6 (two repeat units) geometry was optimized in gas phase at the camb3lyp/6-311g(d,p) level of theory. The trimer systems were constructed by placing the Y6 dimer and PM6 in a pi-pi stacking manner with the intermolecular distances (PM6 to the nearest Y6 molecule) 4.5 Å. We then optimized Y5 single molecule in gas phase at the camb3lyp/6-311g(d,p) level of theory. As the Y5 and Y6 molecules are very similar in their chemical formations, we utilized the Y6+Y6+PM6 trimer geometries, and only replace each Y6 molecule into the optimized Y5 molecules to set up the Y5+Y5+PM6 model trimer system. The Y6+PM6 and Y5+PM6 dimer systems are formed by simply removing the edge Y6 or Y5 molecule respectively, from the corresponding trimer systems.

The adiabatic excited states of all the system are calculated at the omega-tuned camb3lyp/6-311g(d,p) level of theory. We then performed the diabatization analysis for all the systems. The electronic coupling strengths were calculated with the diabatization scheme from ref[5] at the same level of the excited state calculation. Further details on analysis are provided in Supplementary Note 3.

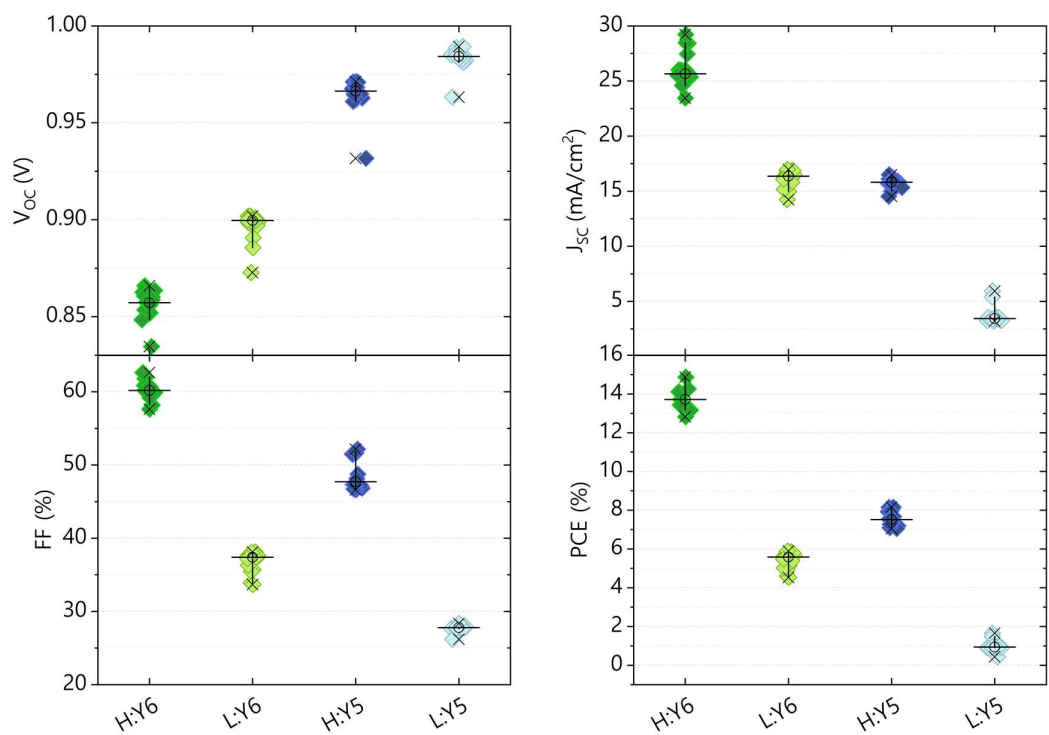

***Figure S1.*** *Statistics of the photovoltaic parameters obtained from current-voltage measurements on OSC devices with high molecular weight PM6 (H) and low molecular weight PM6 (L) with Y5 and Y6 non-fullerene acceptors. The statistics were obtained from 12 to 18 devices with a device are of 0.06 cm², and an active layer thickness of 100-110 nm.*

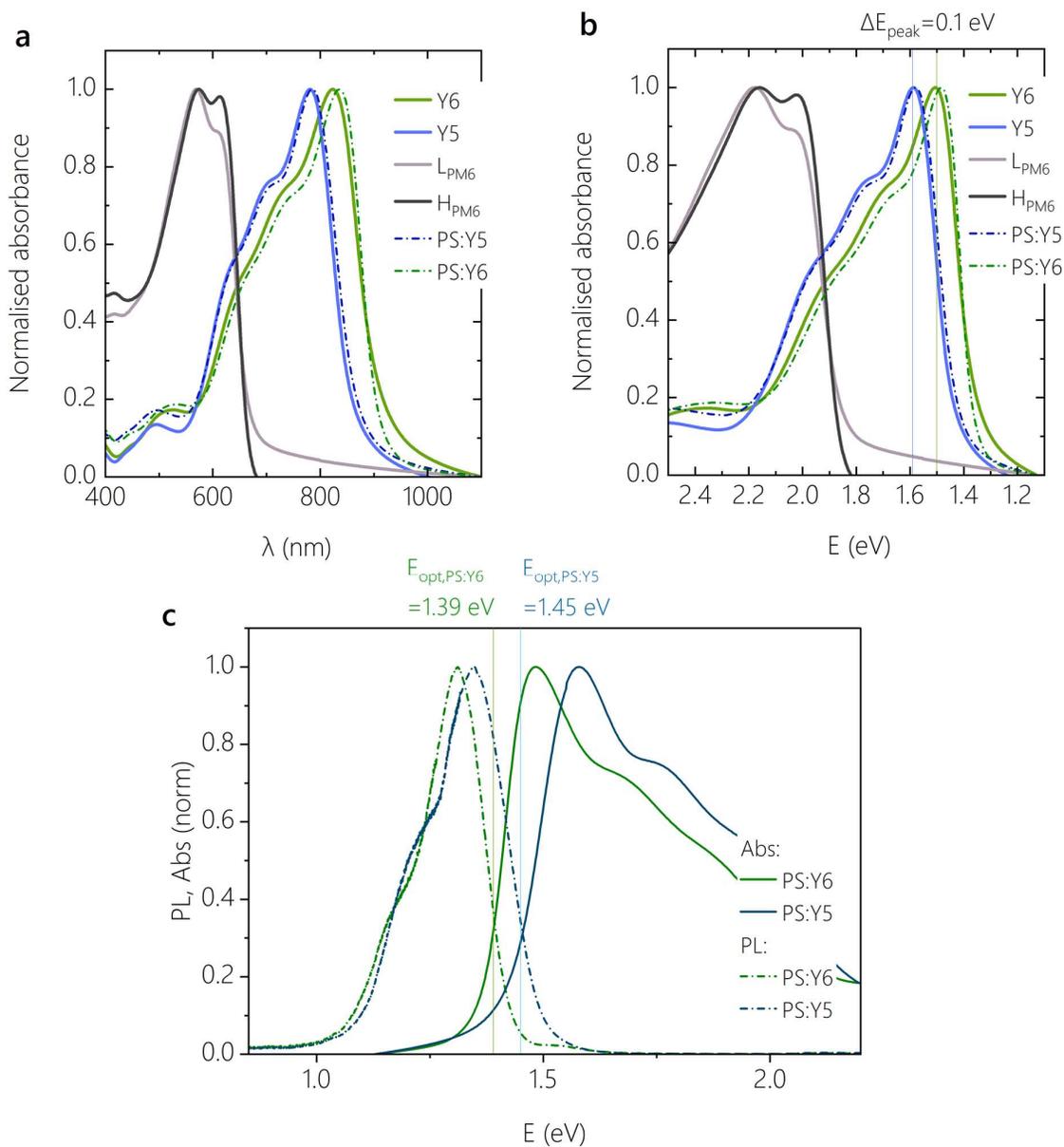

***Figure S2***. *The normalised absorption spectra of the neat films of* $H_{PM6}$, $L_{PM6}$ *and the Yx acceptor components (as neat films and as dispersed in a 1:1.2 ratio in a polystyrene, PS, matrix), plotted as a function of (a) wavelength and (b) photon energy. The reduced polymer aggregation in* $L_{PM6}$ *is seen from a reduced 0-0 to 0-1 peak ratio. (c) Normalised absorption and photoluminescence spectra of the PS:NFA films, representing the higher optical bandgap for Y5.*

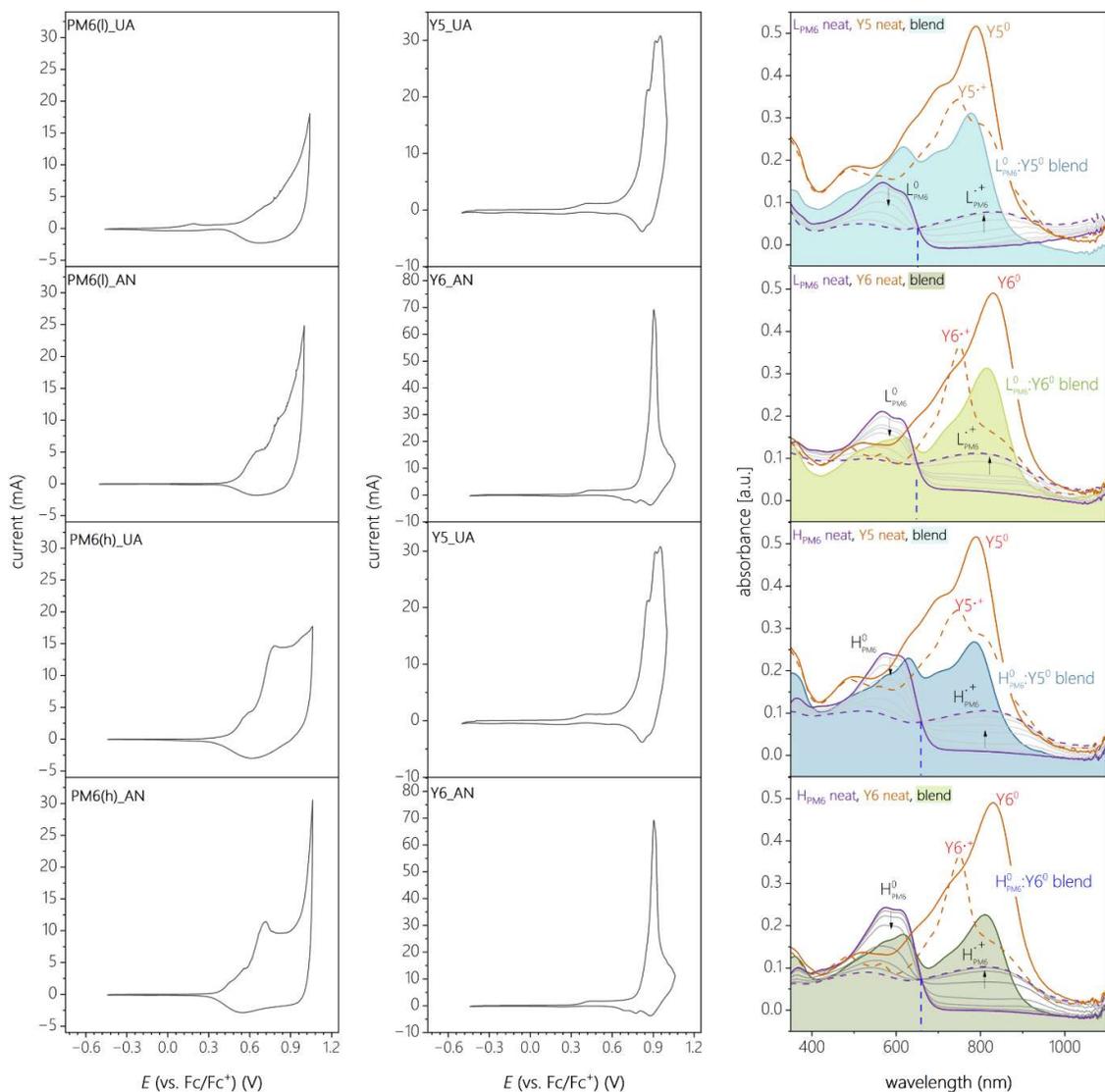

***Figure S3.*** *(left and center) CV of neat films of PM6 polymers of different molecular weight, as well as that of neat Y5 and Y6. (right) Absorption spectra recorded during cyclic voltammetry of the neat polymer and NFA films, and the corresponding blends in neutral form (at 0.00 V vs Fc/Fc+). The dotted lines denote the absorption spectra of the radical cationic species formed at higher potentials on each of the neat components $L_{PM6}$ and Y5.*

| neats | | | |
|---|---|---|---|
| | **N** | **R•+** | **iso** |
| **L$_{PM6}$** | 570 nm / 620 nm | 815 nm | 660 nm |
| **H$_{PM6}$** | 570 nm / 610 nm | 815 nm | 660 nm |
| **Y5** | 787 nm | 751 nm | - |
| **Y6** | 830 nm | 751 nm | - |
| **L:Y5 blend** | | | |
| | **N** | **R•+** | **iso** |
| **PM6** | 620 nm | 650 nm - 780 nm | 650 nm |
| **Y5** | 780 nm | 751 nm | - |
| **L:Y6 blend** | | | |
| | **N** | **R•+** | **iso** |
| **PM6** | 610 nm | 650 nm - 814 nm | 650 nm |
| **Y6** | 814 nm | 751 nm | - |
| **H:Y5 blend** | | | |
| | **N** | **R•+** | **iso** |
| **PM6** | 630 nm | 650 nm - 785 nm | 660 nm |
| **Y5** | 780 nm | 751 nm | - |
| **H:Y6 blend** | | | |
| | **N** | **R•+** | **iso** |
| **PM6** | 620 nm | 650 nm - 814 nm | 660 nm |
| **Y5** | 814 nm | 751 nm | - |

*Figure S4.* A view of the spectral positions of characteristic bands of neutral and radical cation species of the donor and acceptors, as measured in neat films as well as blend films. Also highlighted is the isosbestic point of the polymer spectral evolution.

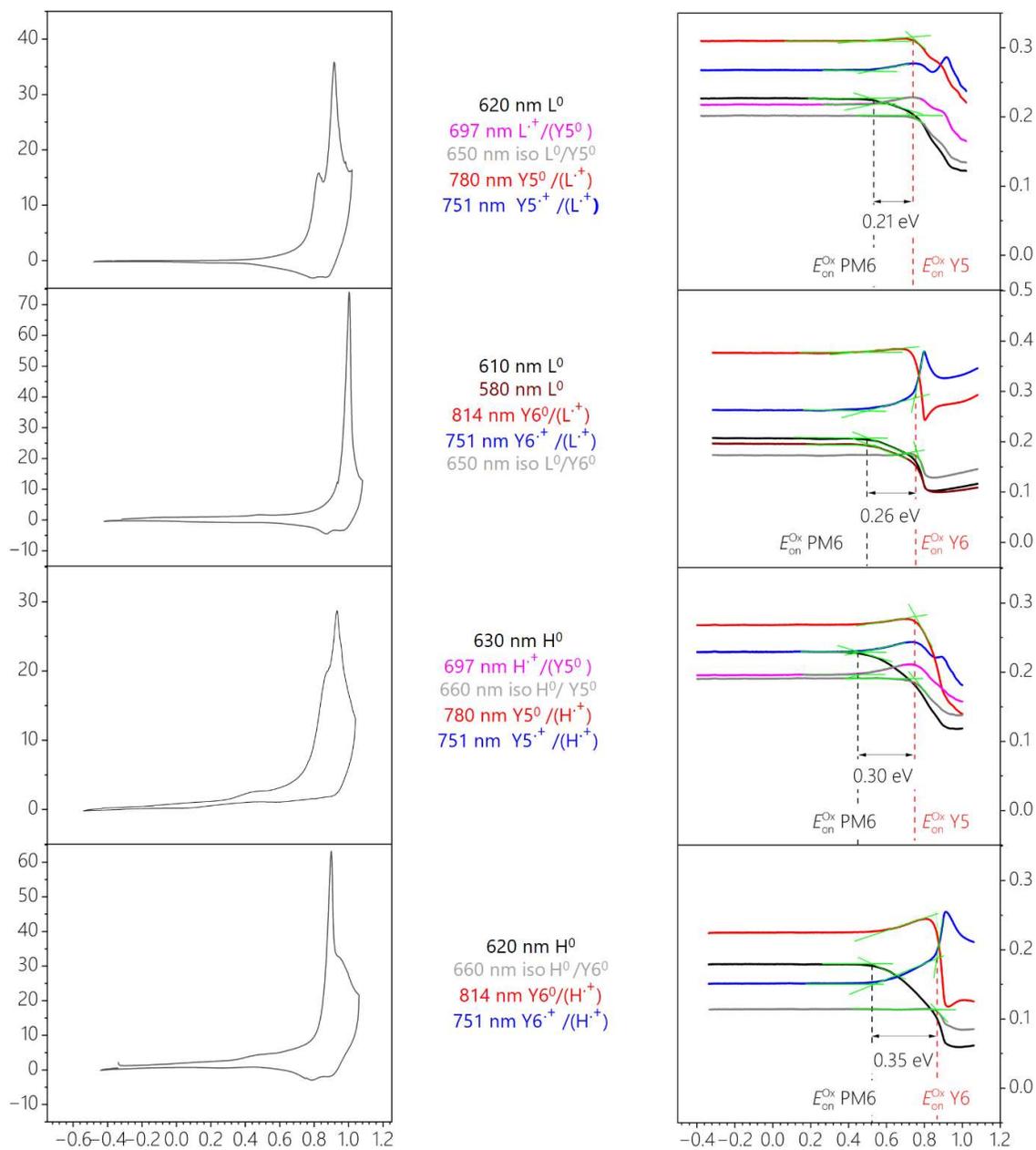

*Figure S5.* (left panels) In-situ CV plots measured on the four model blends: top-to-bottom L:Y5, L:Y6, H:Y5, H:Y6. (right panels) corresponding peak trend for the different blends L:Y5, L:Y6, H:Y5 and H:Y6 (top to down). The spectral evolution of the donor and acceptor band peaks are fitted with the tangent method to obtain the oxidation onsets of the blend constituents as measure within the blend, whose difference yields the ionization energy between the blend constituents.

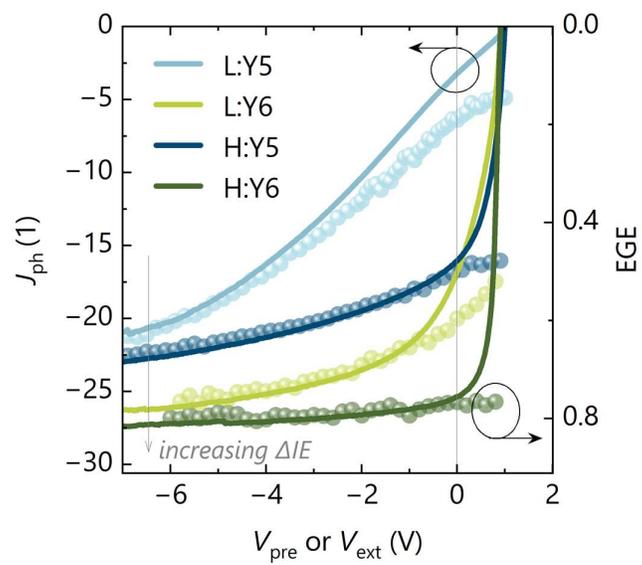

*Figure S6.* High reverse bias overlay of photocurrent (*J*ph) from JV characteristics and external free charge generation efficiency (EGE) from modified time delayed collection field (mTDCF) measurements.

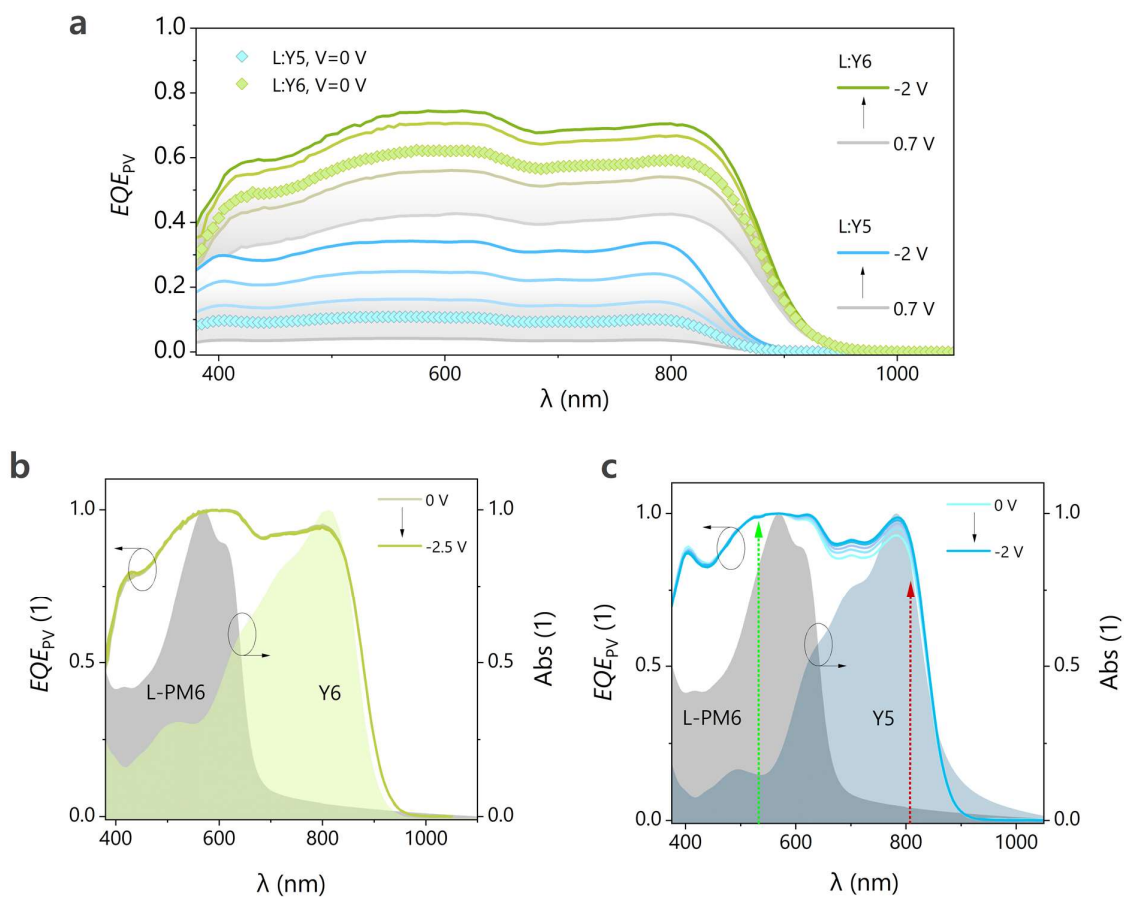

***Figure S7.*** *(a) EQE$_{PV}$ as a function of bias for L:Y6 (in green) and L:Y5 (in blue). (b) Normalized EQE$_{PV}$ as a function of bias for L:Y6, showing that the shape of the spectrum is independent of bias. The right y-axis is the normalised absorption, to show the spectral regions of absorption dictated by the donor and acceptor as indicated by the shaded portions. (c) Normalized EQE$_{PV}$ as a function of bias for L:Y5, wherein the marginal enhancement of the Y5-relevant peak can be observed. The vertical coloured arrows indicated the two laser wavelengths used to selectively excite the L:Y5 system in excitation-dependent TDCF measurements, shown in Figure S6.*

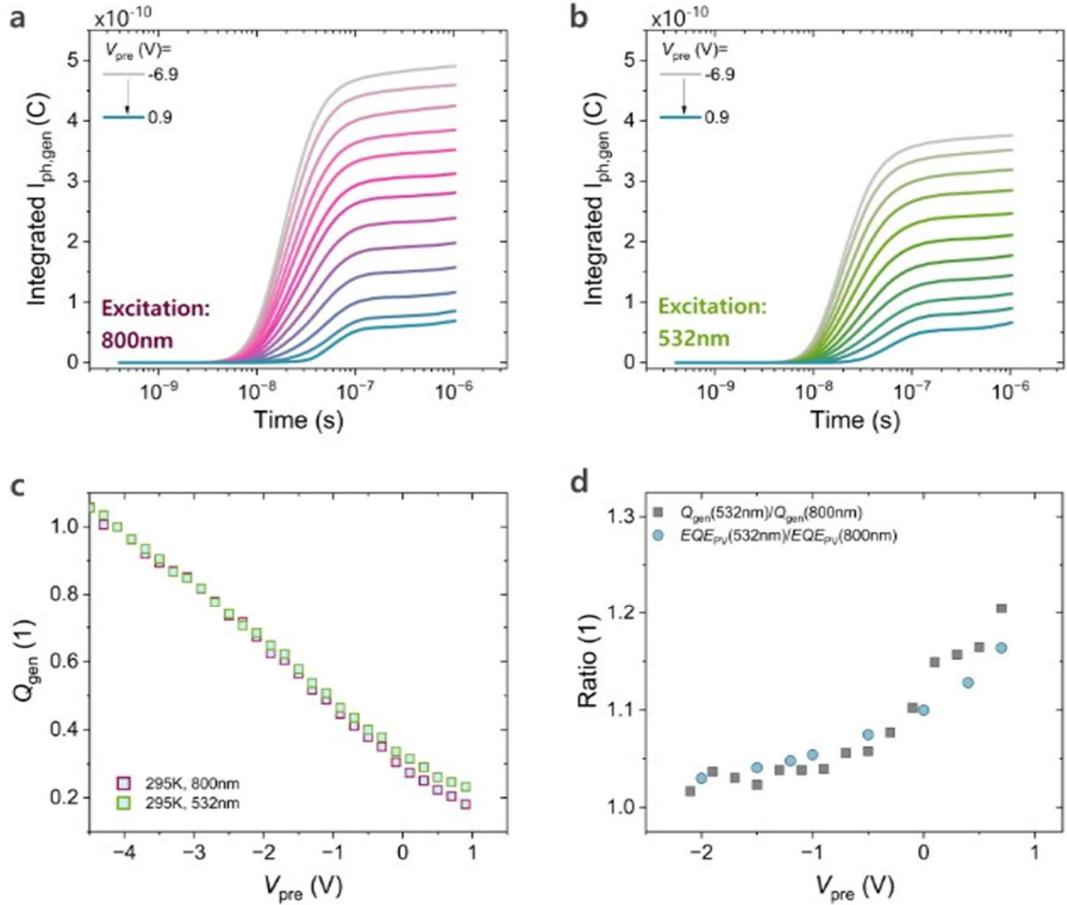

***Figure S8.*** *(a-b) Temporal evolution of generated charge from the integrated transient photocurrent, recorded from TDCF measurements, at different pre-bias conditions ($V_{pre}$) for the L:Y5 system, for (a) λ=800nm excitation and (b) λ=532nm excitation. (c) Normalized free charge plotted normalised at $V_{pre}$=-4 V. The bias-dependence of free charge generation is identical for the two selective excitations, with marginal divergence occurring in the more positive voltage range. In this range, excitation of the acceptor yields marginally more field-dependence than for donor excitation, in accordance with $EQE_{PV}$ data, with the difference in bias-dependences being less than 5%. (e) The ratio of free charge as well as the $EQE_{PV}$ for selective excitations, plotted as a function of the applied bias to the device ($V_{pre}$ in case of free charge), showing agreement between the field-dependences recorded from both techniques for the L:Y5 system.*

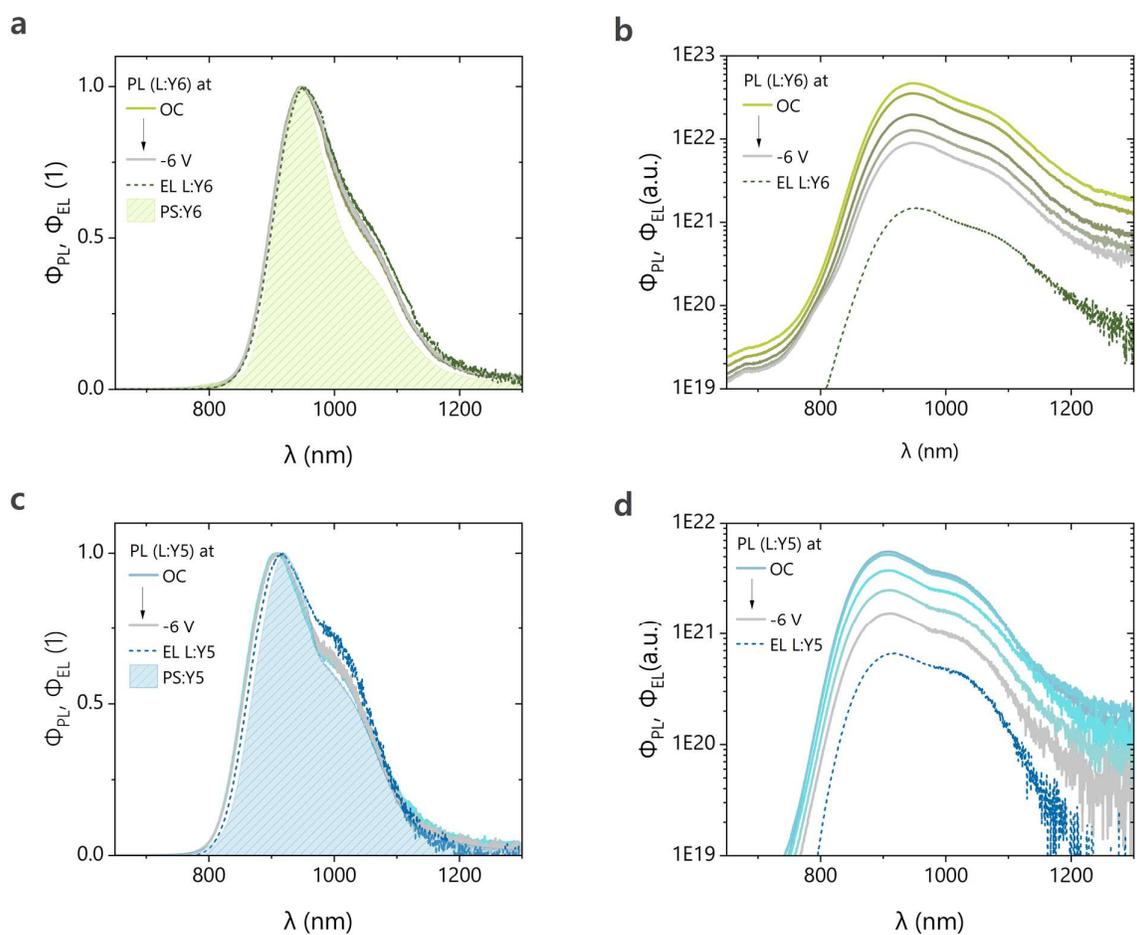

***Figure S9.*** *(a) Normalised EL and bias-dependent PL of a L:Y6 blend device, compared to a PL spectrum for a polystyrene:Y6 blend device at OC. (b) EL and bias-dependent PL spectra of L:Y6 plotted on a semi-log scale. (c) The equivalent of subgraph a, but for L:Y5. (d) The equivalent of subgraph b, but for L:Y5. The overlap of all spectra in subgraphs a and c shows that the radiative decay pathway for free charges too occurs via the NFA singlet in these systems, and that the contribution of CT emission in the PL and EL spectra of the device is likely negligible.*

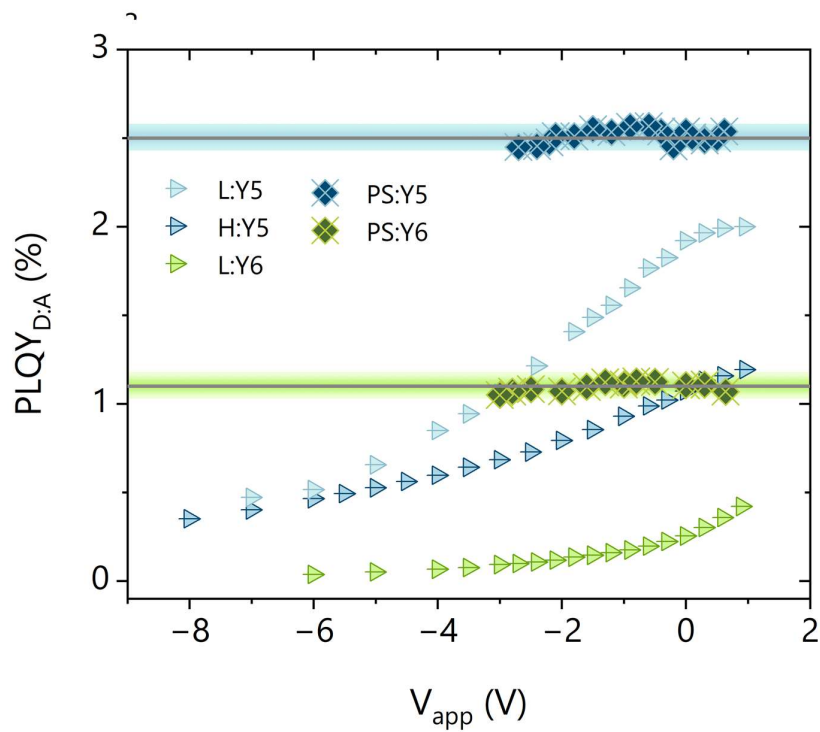

***Figure S10.*** *Experimental PLQY values extrapolated as a function of external bias using the ssPL photon flux, for the inefficient blend systems as well as polystyrene:NFA devices in the same blend ratio (1:1.2). The conversion was done using the equation* $PLQY_{D:A}(V) = \frac{\phi_{PL,max}(V)}{\phi_{PL,max}(V_{OC})} \cdot PLQY_{D:A}(V_{OC})$. *The emission from blends is clearly bias-dependent, while that in blends of the NFA in a polystyrene matrix are bias-invariant. This means that the bias-dependent process impacting emission from the blend occurs at the sites of donor acceptor interaction.*

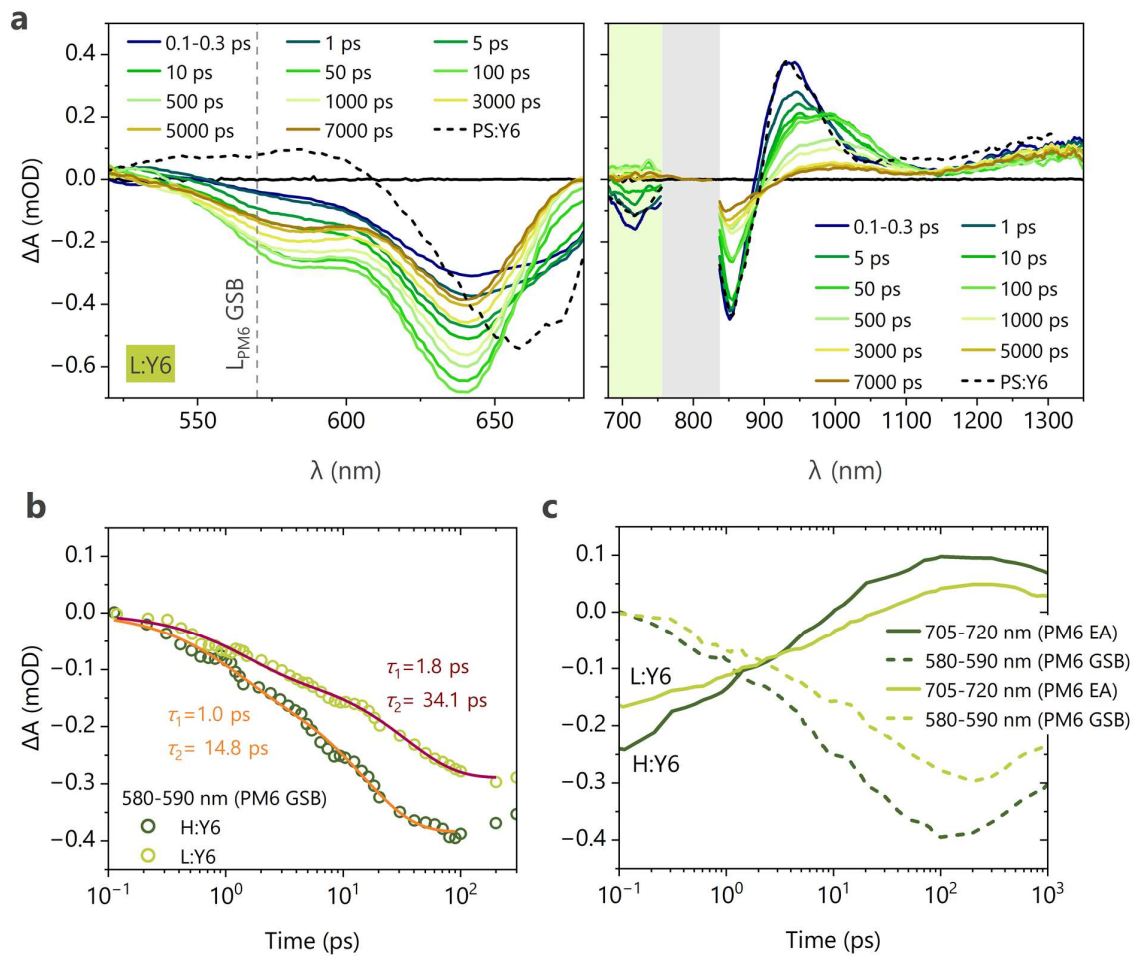

***Figure S11.*** *(a) Transient absorption spectra in the visible and infrared region, probed on an L:Y6 film of excited with a 1.77eV laser pulse of 2 μ/cm² fluence for preferential NFA excitation. The grey shaded area denotes the region of optical excitation of the probe beam. (b) Comparison of the PM6 ground state bleach (GSB) dynamics for H:Y6 and L:Y6 blends, showing a faster rise in the case of H:Y6 with higher ΔIE. (c) System-wise comparison of both PM6 GSB dynamics and PM6 electroabsorption (EA) signals corresponding to charge formation.*

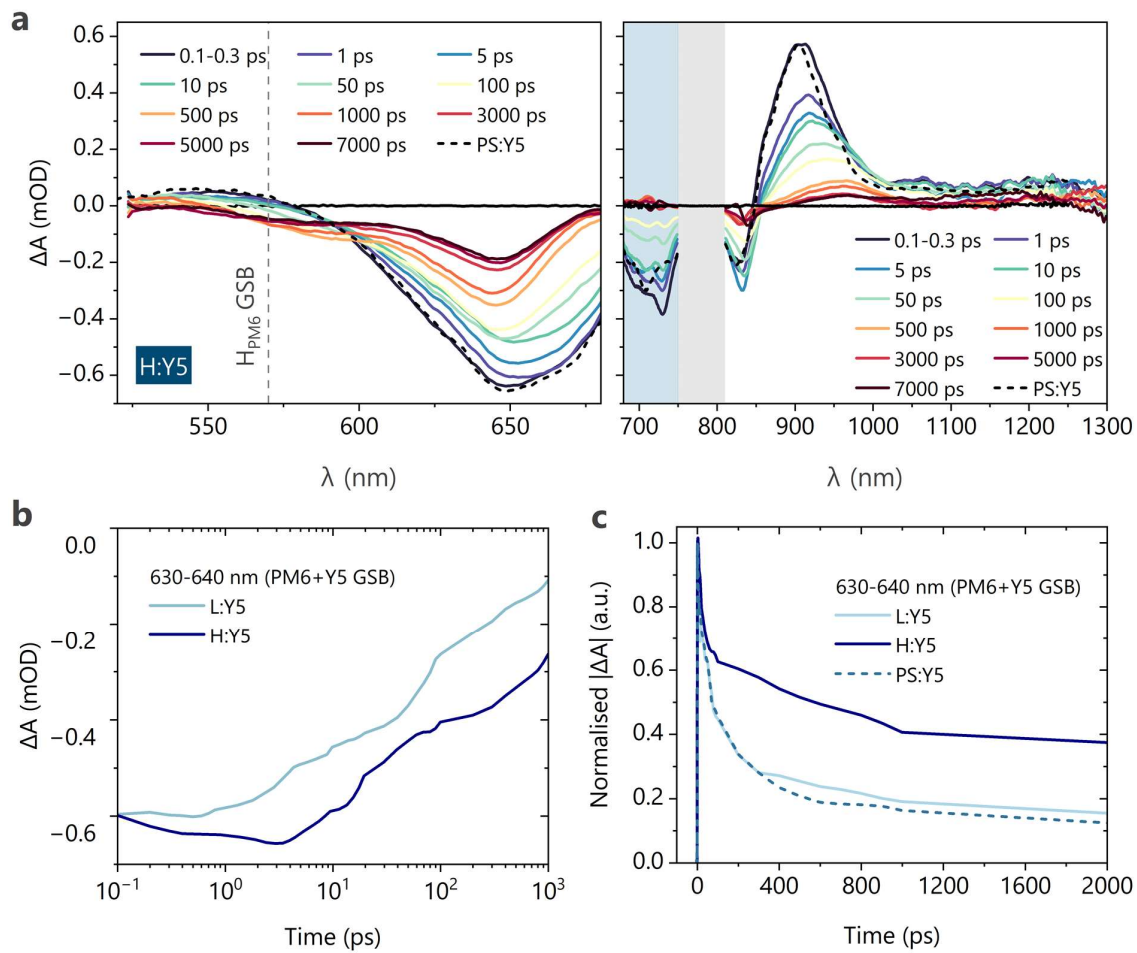

*Figure S12.* (a) Transient absorption spectra in the visible and infrared region, probed on an H:Y5 film of excited with a 1.77eV laser pulse of 2 μJ/cm² fluence for preferential NFA excitation. The grey shaded area denotes the region of optical excitation of the probe beam. (b) Comparison of the PM6 and Y5 ground state bleach (GSB) dynamics for H:Y5 and L:Y5 blends, showing a rise in the case of H:Y5 with higher ΔIE. (c) Longer time dynamics of the PM6+Y5 GSB features in Y5-based blends, normalised to the initial intensity of the two blends along with the normalised GSB decay of PS:Y5.

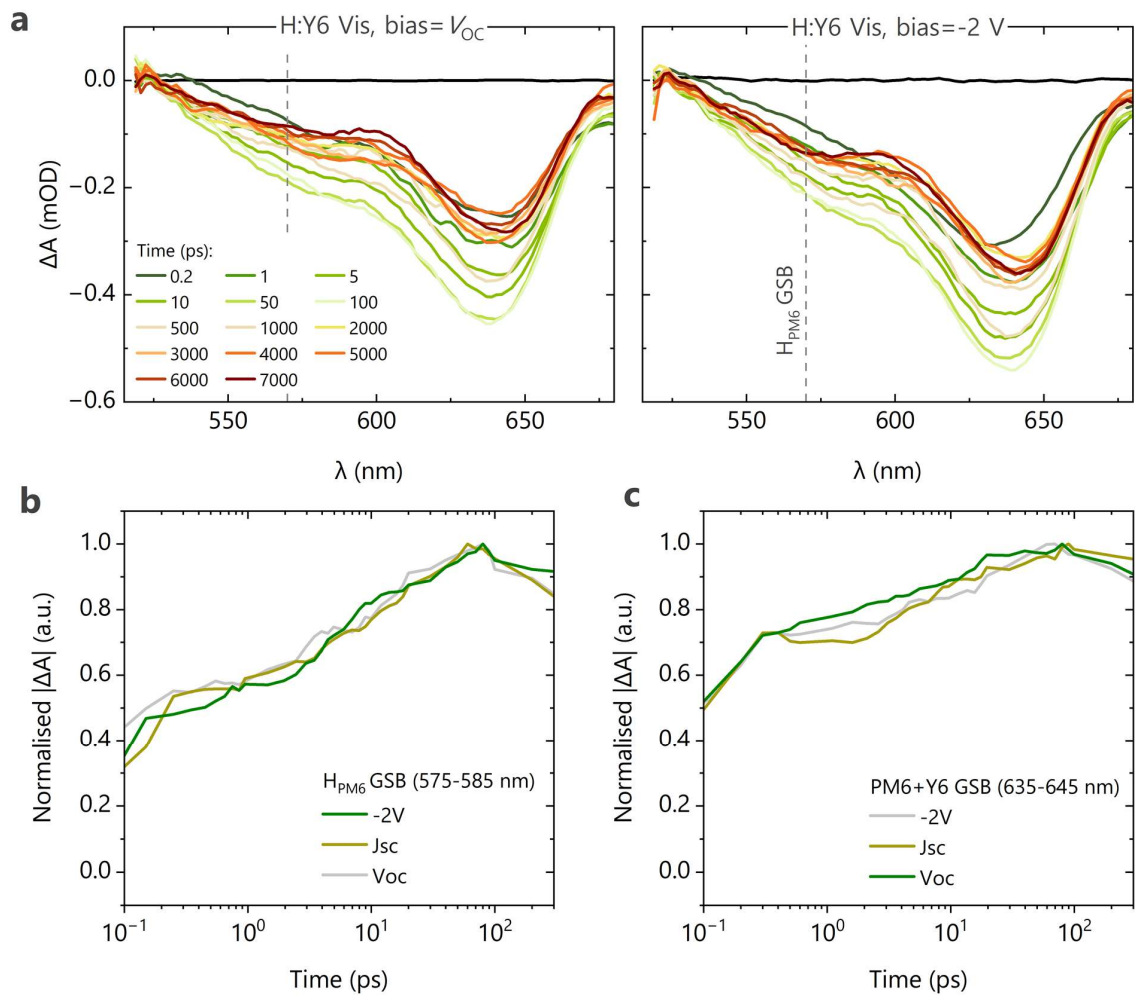

*Figure S13. (a) Bias-dependent transient absorption spectra in the visible region probed on an H:Y5 device, when biased externally with $V_{OC}$ and -2 V. The semi-transparent sample was excited with a 1.77eV laser pulse of 7 µJ/cm² fluence for preferential NFA excitation. (b) Bias-dependent TA dynamics of the PM6 relevant GSB band in H:Y6 devices, showing no alternation of the free charge generation properties with an external field. (c) The equivalent of subgraph b, but of the GSB band around 640 nm relevant to both PM6 and Y6.*

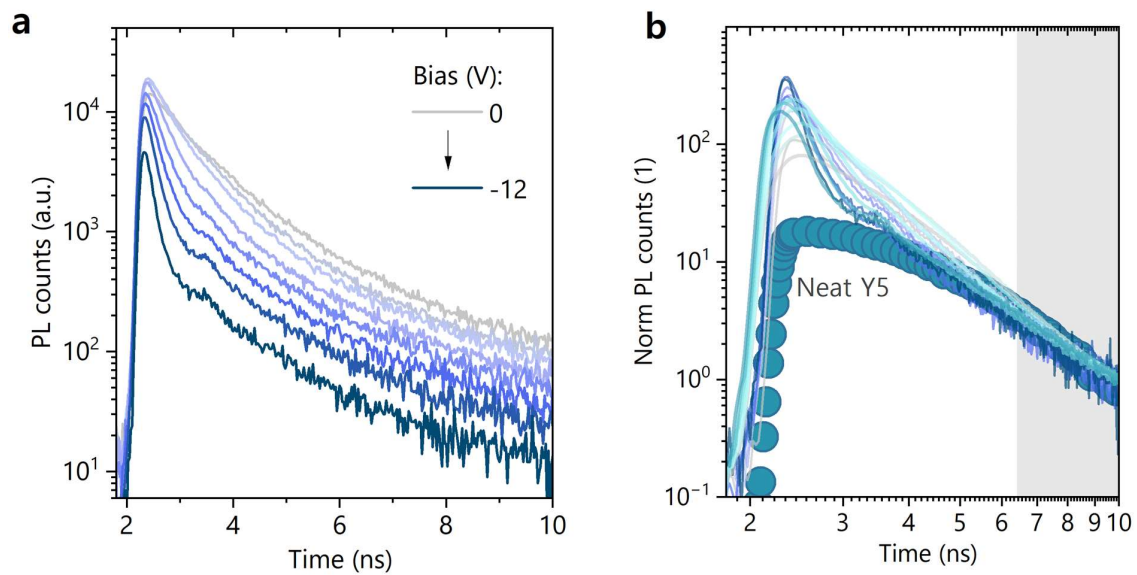

*Figure S14.* (a) TrPL decays for the H:Y5 devic. (b) Normalised TrPL kinetics of H:Y5 and L:Y5 devices, as a function of external bias. The intensities at each bias are normalised to the respective values at 10 ns. Overlaid on top is the TrPL kinetics of PS:Y5, indicating the emission from undissociated LE excitons in Y5 domains at later times in both blends.

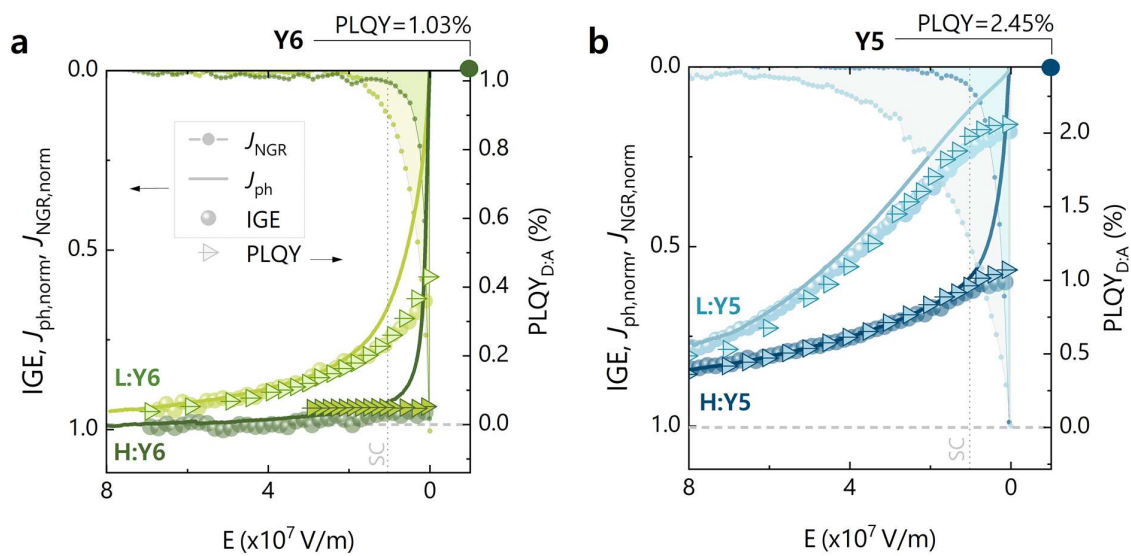

***Figure S15.*** *(a) An overlay of $J_{ph}$, EGE, $J_{NGR}$, and PLQY as a function of the effective electric field across the active layer for the Y6-based model systems. An excellent anticorrelation of the generation and emission properties is observed across the entire effective field range. (b) The equivalent of subgraph a, but for the Y5-based blends. The vertical dotted line marks the effective electric field at short-circuit conditions. Importantly, the scenario of zero free charge generation (open-circuit) corresponds to emission that equals the PLQY of the respective NFA. Hence, the anticorrelation of emission and generation dictates that the PLQY from the blend at a certain bias be represented as a fraction of the PLQY of the neat NFA, depending on the efficiency of free charge generation at that bias condition.*

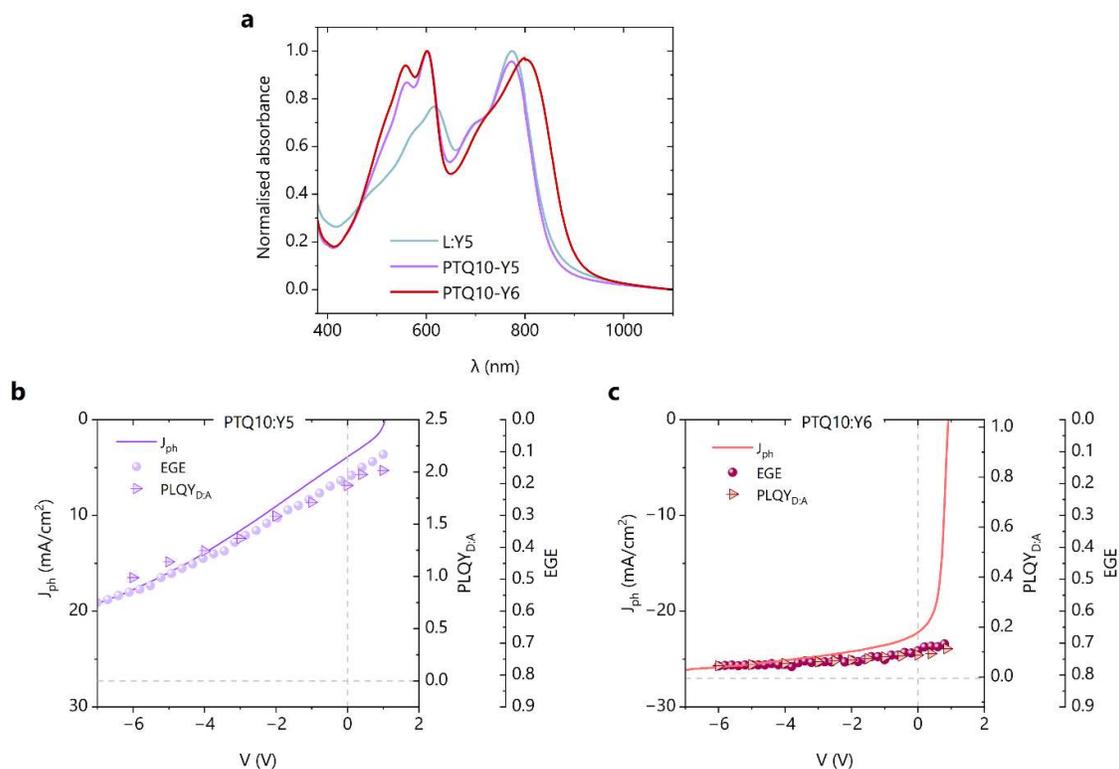

***Figure S16.*** *(a) Normalised absorption spectra for the PTQ10:Yx systems. Overlaid is the normalised absorbance of L:Y5 for comparison. (b-c) An overlay of $J_{ph}$ from JV characteristics, EGE from mTDCF and $PLQY_{D:A}$ from ssPL data plotted for the PTQ10:Y5 and PTQ10:Y6 OSC, respectively.*

|  | L:Y5 | PTQ10:Y6 | PTQ10:Y5 |
|---|---|---|---|
| Voc (V) | 0,99 | 0,88 | 1,00 |
| Jsc (mA/cm²) | 3,48 | 23,1 | 4,3 |
| FF (%) | 27,65 | 59,54 | 37 |
| PCE (%) | 0,96 | 12,64 | 1,91 |

***Table S1.*** *Photovoltaic parameters for the PTQ10 polymer based OSCs systems, along with the parameters of L:Y5 for comparison.*

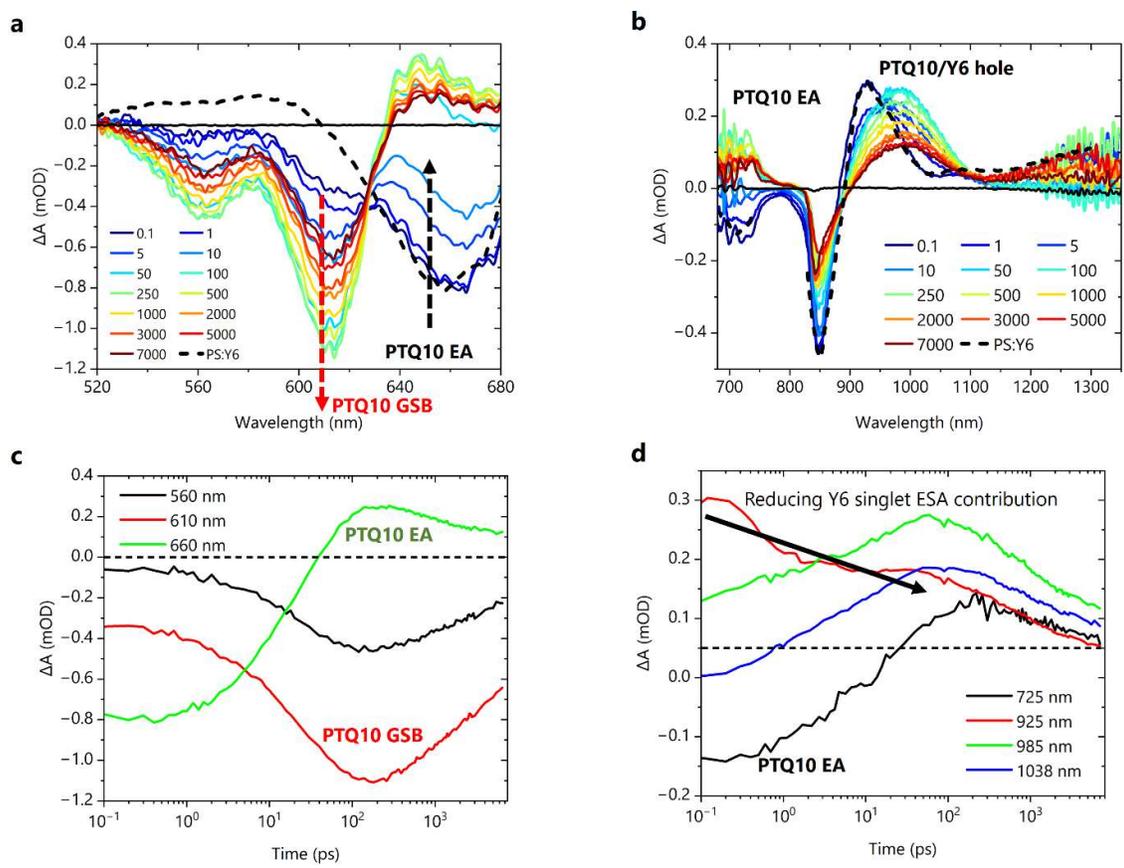

*Figure S17.* (a-b) Transient absorption spectra in the visible and infrared regions, probed on a PTQ10:Y6 film of excited with a 1.6eV laser pulse of 2 μJ/cm² fluence for preferential NFA excitation. (c-d) Kinetic comparison of TA spectral features in the visible region and NIR region, respectively.

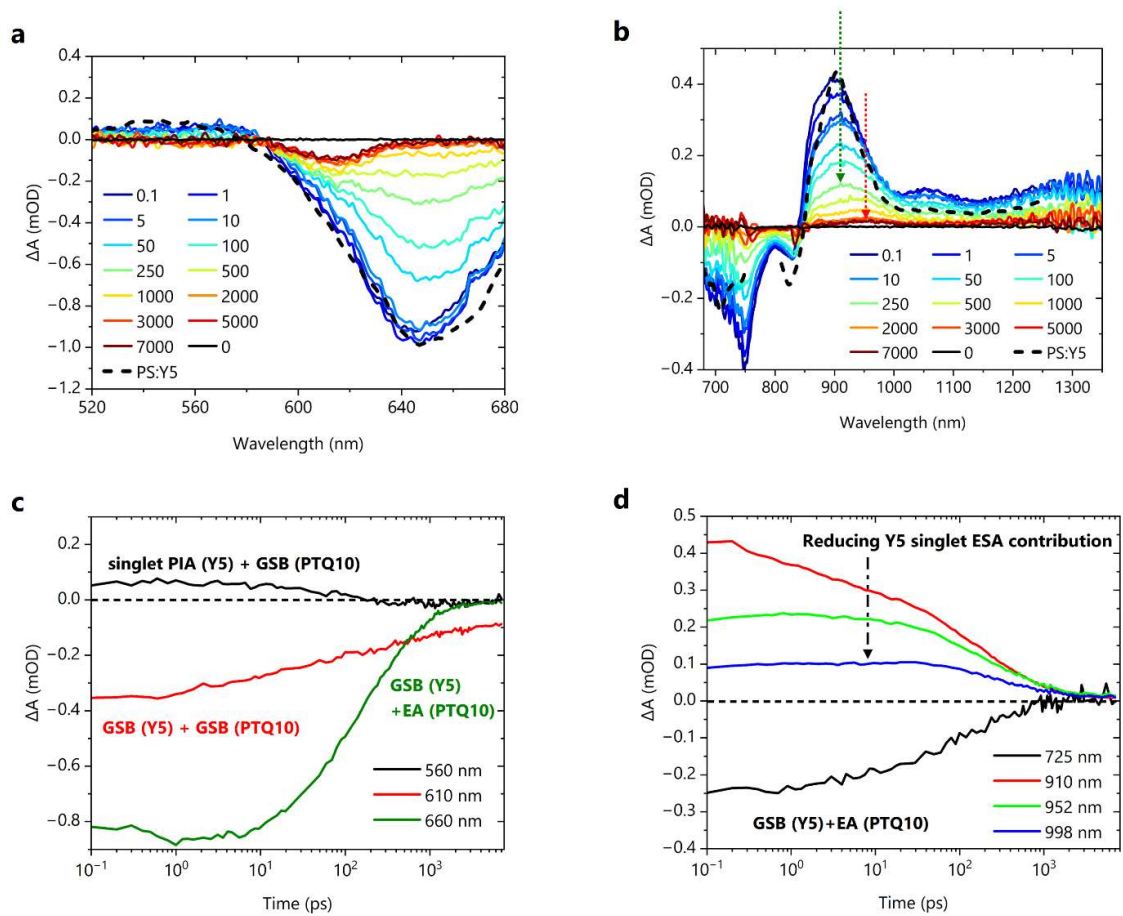

*Figure S18.* (a-b) Transient absorption spectra in the visible and infrared regions, probed on a PTQ10:Y5 film of excited with a 1.6eV laser pulse of 2 μJ/cm² fluence for preferential NFA excitation. (c-d) Kinetic comparison of TA spectral features in the visible region and NIR region, respectively.

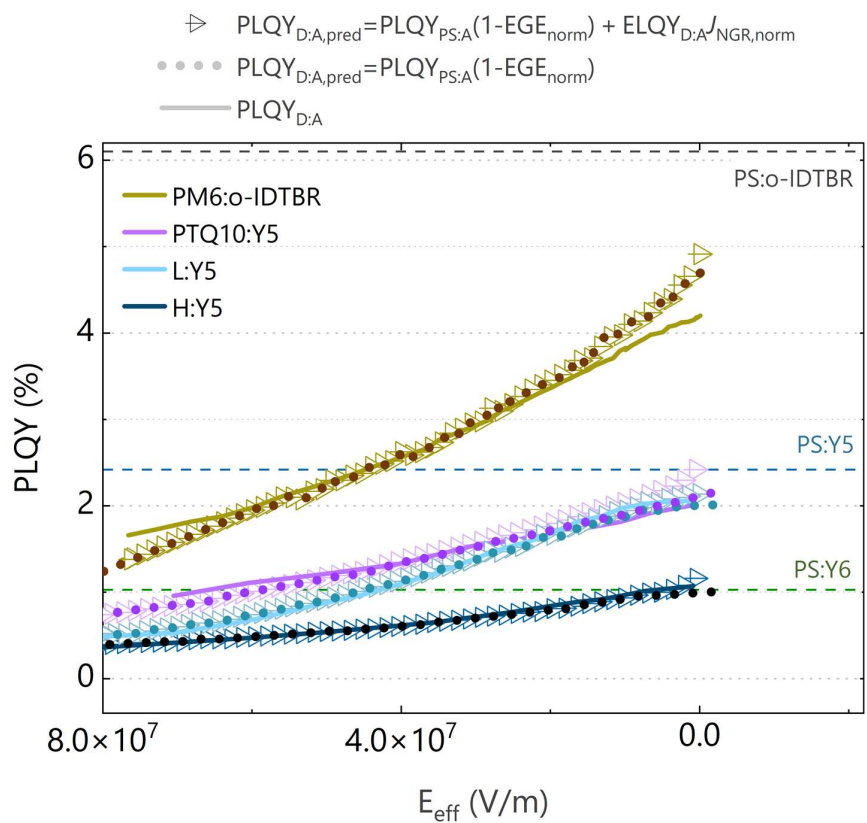

*Figure S19.* Overlay and comparison of the field-dependence of the theoretical predicted $PLQY_{D:A,pred}$ (triangles) reconstructed using equation (2) in the main text, and the experimental $PLQY_{D:A}$ (solid lines) obtained from field-dependent ssPL and equation (1) in the main text. In addition, the predicted $PLQY_{D:A,pred}$ without a non-geminate recombination component is plotted (dotted).

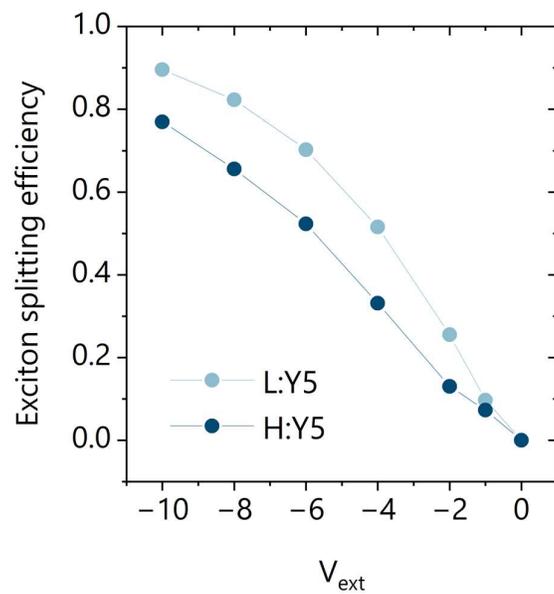

*Figure S20.* Bias-dependence of exciton splitting efficiency for the L:Y5 and H:Y5 devices, relative to the emission at 0 V, obtained by integrating the TrPL kinetics of each blend.

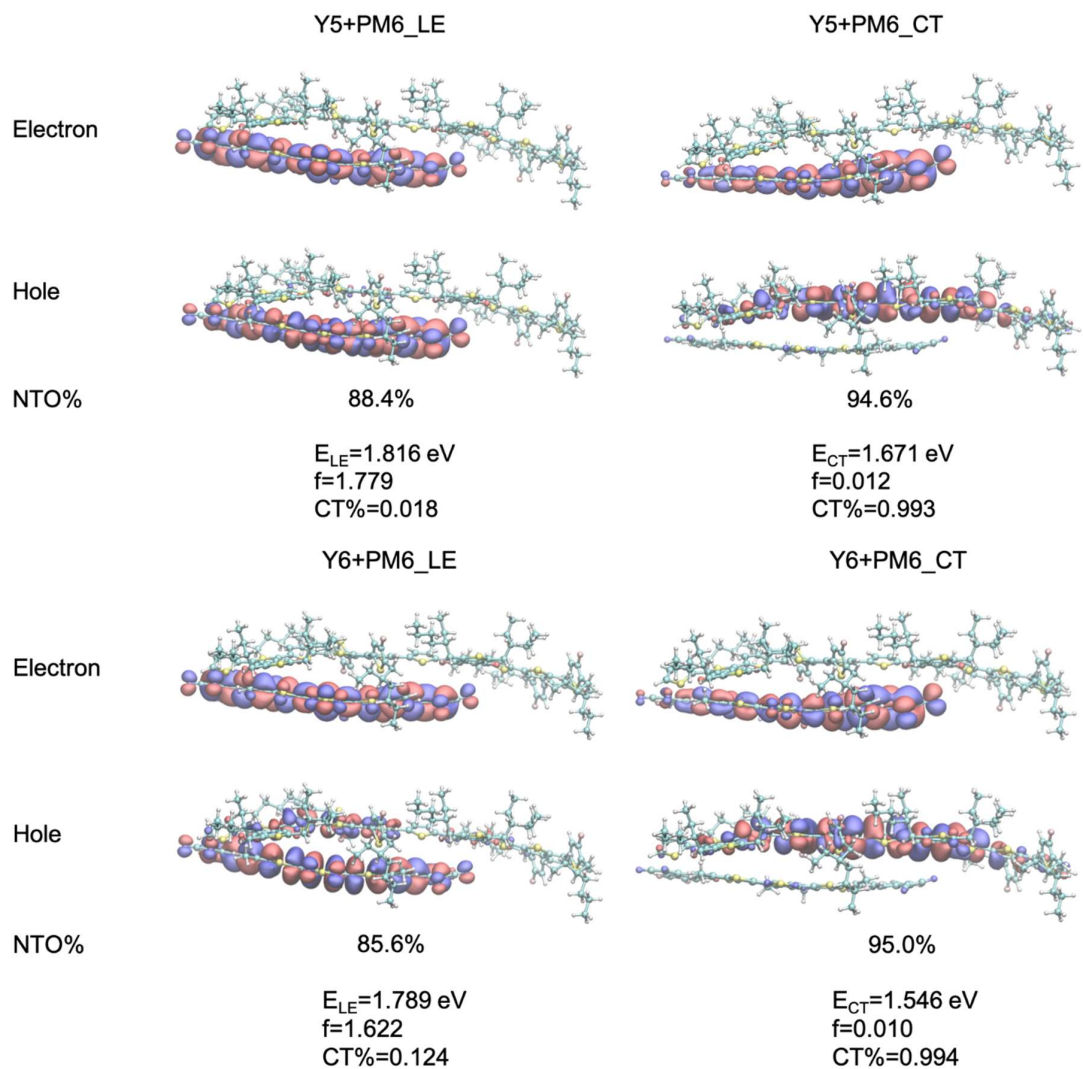

|  | Y5+PM6_LE | Y5+PM6_CT |
|---|---|---|
| Electron | | |
| Hole | | |
| NTO% | 88.4% | 94.6% |

$E_{LE}$=1.816 eV
f=1.779
CT%=0.018

$E_{CT}$=1.671 eV
f=0.012
CT%=0.993

|  | Y6+PM6_LE | Y6+PM6_CT |
|---|---|---|
| Electron | | |
| Hole | | |
| NTO% | 85.6% | 95.0% |

$E_{LE}$=1.789 eV
f=1.622
CT%=0.124

$E_{CT}$=1.546 eV
f=0.010
CT%=0.994

*Figure S21.* Natural transition orbitals (hole and electron) of the singlet local excited (LE) and charge transfer (CT) state of the Y5+PM6 and Y6+PM6 system, together with the corresponding weights (NTO%). The excitation energies ($E_{LE}/E_{CT}$), oscillator strength (f) and weight of CT character are summarized.

**Table S2.** *Energies of excited state energies (local singlet state, LE, and charge-transfer state, CT) in interfacial dimer (Yx+PM6) and interfacial trimer (2Yx+PM6) systems, used in computational modelling. The difference in LE-CT state energies, as well differences in the energies in Y5- and Y6-based systems are also provided.*

| Energy of: | CT | LE | CT | LE |
|---|---|---|---|---|
| Interfacial system: | 2Y5+PM6 | 2Y5+PM6 | Y5+PM6 | Y5+PM6 |
| @ 4.5 Å | 1.696 | 1.790 | 1.671 | 1.816 |
| LE-CT for Y5 | 0.095 | | 0.145 | |
| | | | | |
| | CT | LE | CT | LE |
| | 2Y6+PM6 | 2Y6+PM6 | Y6+PM6 | Y6+PM6 |
| @ 4.5 Å | 1.550 | 1.770 | 1.546 | 1.789 |
| LE-CT for Y6 | 0.220 | | 0.243 | |
| | | | | |
| Y5-Y6 | 0.146 | 0.020 | 0.125 | 0.027 |

**Table S3.** *Energies (in Hartree) used in the calculation of forward and backward reorganisation energies, used for determining the inner reorganisation energy necessary for the steady-state rate model in the main text.*

| States | Energy (in Hartree) |
|---|---|
| PM6 | |
| $E_{nC}$ | -12423.6577196 |
| $E_{nN}$ | -12423.6664638 |
| $E_{cN}$ | -12423.4233529 |
| $E_{cC}$ | -12423.4331599 |
| | |
| Y6 | |
| $E_{eA}$ | -4817.93566447 |
| $E_{eE}$ | -4817.93850848 |
| $E_{aE}$ | -4818.12018875 |
| $E_{aA}$ | -4818.12309981 |
| | |
| Y5 | |
| $E_{eA}$ | -4420.96453717 |
| $E_{eE}$ | -4420.96721174 |
| $E_{aE}$ | -4421.14262068 |
| $E_{aA}$ | -4421.14537882 |

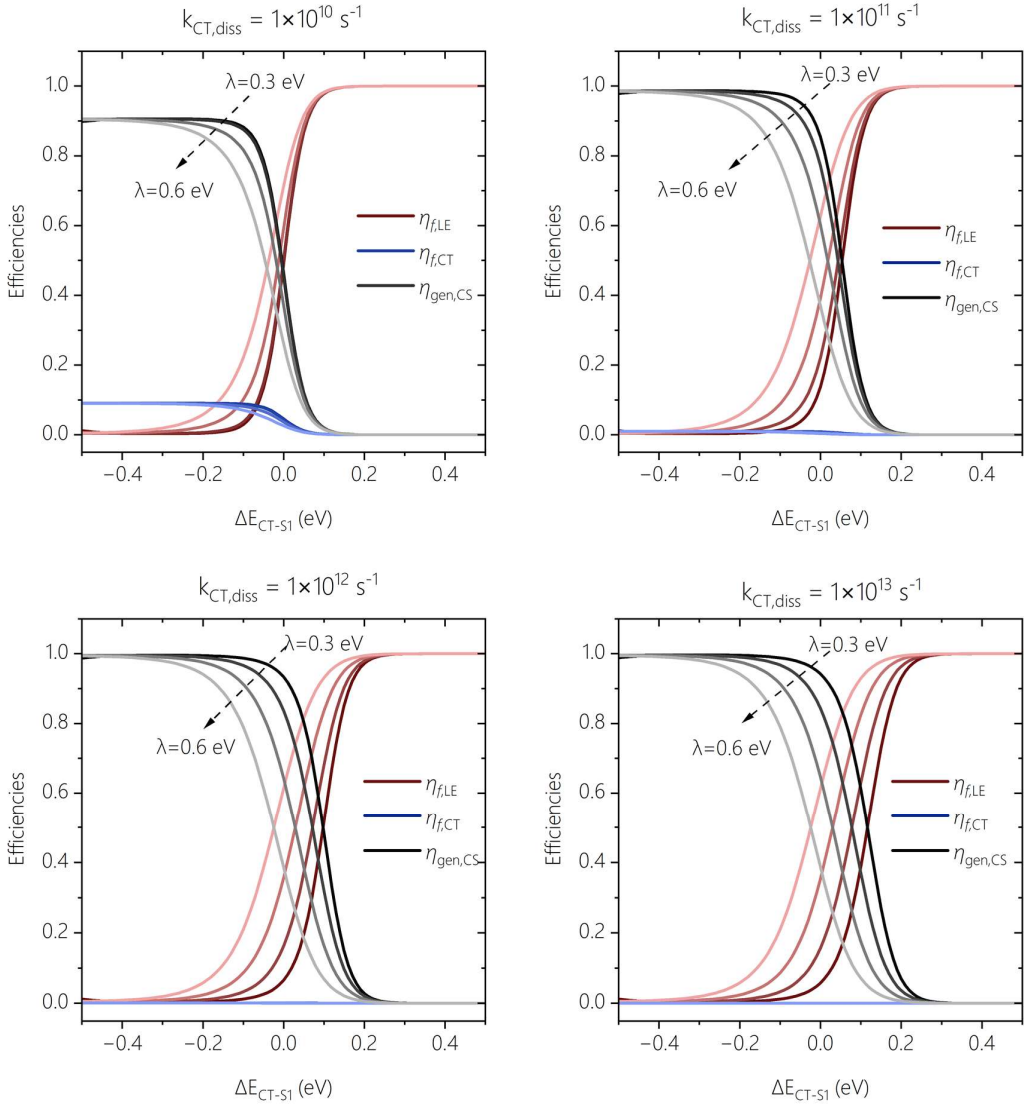

*Figure S22.* The dependence of charge generation efficiency, losses via the decay of the local NFA exciton and the CT state ($\eta_{gen,CS}$, $\eta_{f,LE}$, $\eta_{f,CT}$, respectively) on $\Delta E_{CT-S1}$, simulated from the steady-state rate model assuming not all photogenerated excitons are able to undergo charge transfer (reduced $k_{diss,LE}$). The simulated data is shown for varying reorganisation energies $\lambda$, simulated with the parameters in the given table.

| With reduced $k_{diss,LE}$ (prefactor=0.1 in eq.4a in main text) | |
|---|---|
| Parameter | Value |
| $k_{f,LE}$ and $k_{f,CT}$ | $1 \times 10^9$ s$^{-1}$ |
| $k_{diss,CT}$ | $1 \times 10^{10}$ s$^{-1}$ to $1 \times 10^{13}$ s$^{-1}$ |
| $G$ | $1 \times 10^{28}$ m$^{-3}$s$^{-1}$ |
| $|H_{DA}|$ | 0.01 eV |
| $\lambda$ | 0.3, 0.4, 0.5, 0.6 eV |

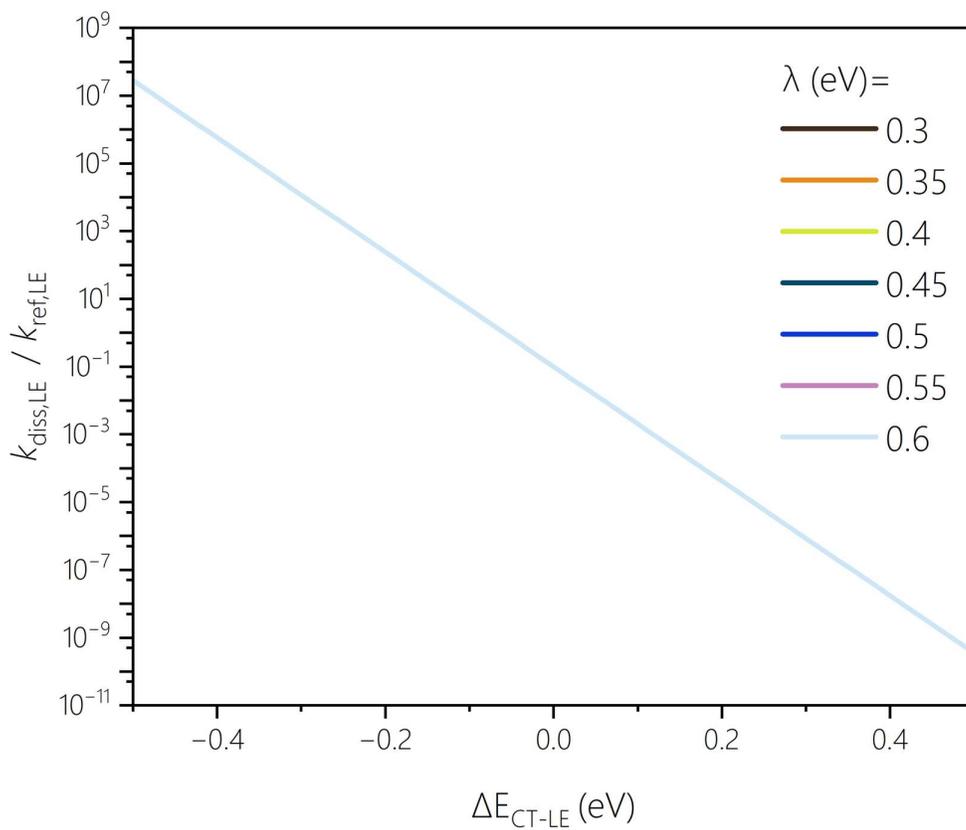

*Figure S23.* The ratio of LE dissociation coefficient ($k_{diss,LE}$) to the LE reformation coefficient ($k_{ref,LE}$), whose values are shown in Figure 7d of the main text, as a function of the CT-LE offset and for various reorganisation energies. The ratio is completely independent of the reorganisation energy.

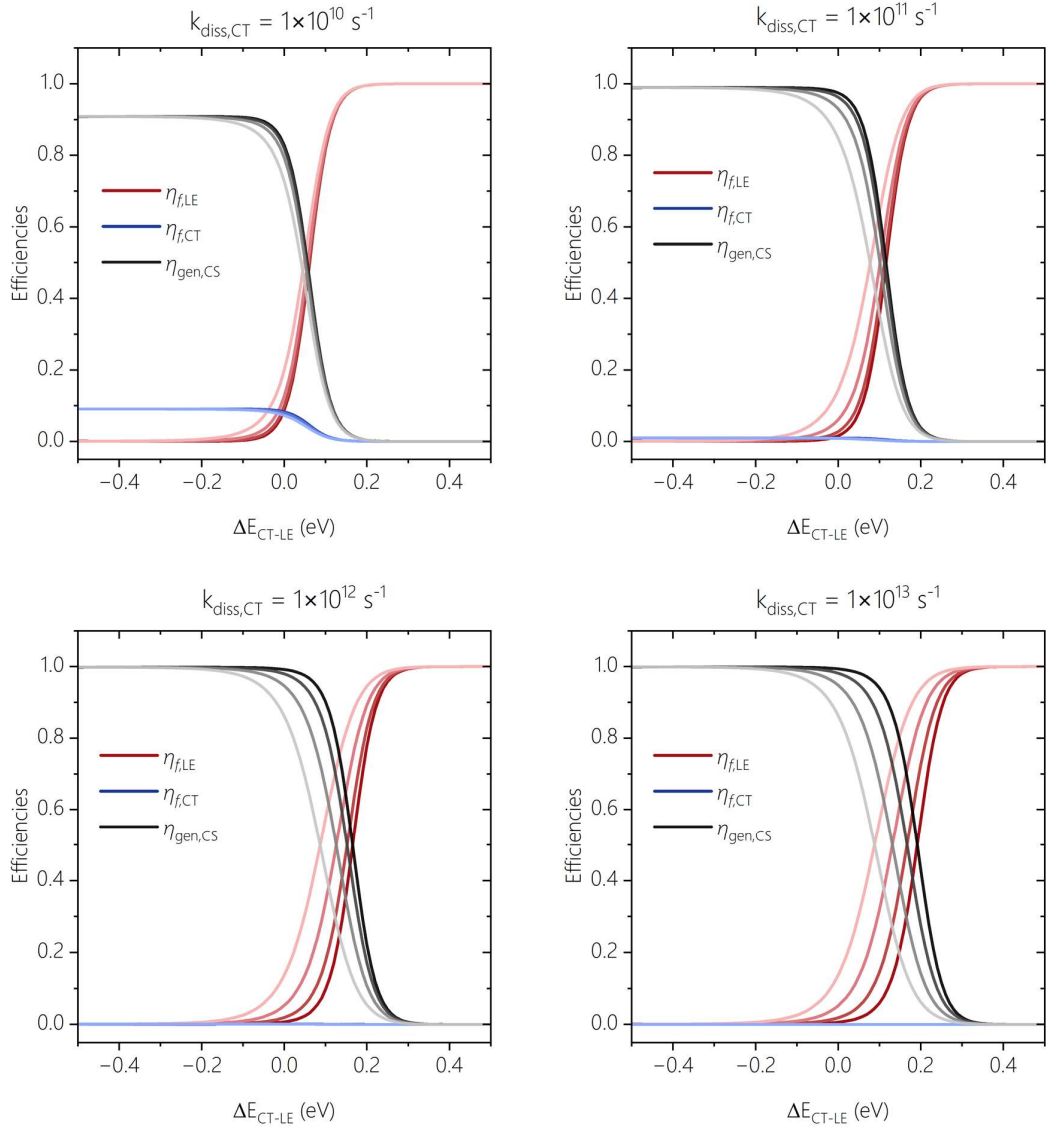

**Figure S24.** *The dependence of charge generation efficiency, losses via the decay of the local NFA exciton and the CT state ($\eta_{gen,CS}$, $\eta_{f,LE}$, $\eta_{f,CT}$, respectively) on $\Delta E_{CT-S1}$, simulated from the steady-state rate model (without a reduction in $k_{diss,LE}$). The simulated data is shown for varying reorganisation energies $\lambda$, simulated with the parameters in the given table.*

| Without reduced $k_{diss,LE}$ (prefactor=1 in eq.4a in main text) | |
|---|---|
| Parameter | Value |
| $k_{f,LE}$ and $k_{f,CT}$ | $1 \times 10^9$ s⁻¹ |
| $k_{diss,CT}$ | $1 \times 10^{10}$ s⁻¹ to $1 \times 10^{13}$ s⁻¹ |
| $G$ | $1 \times 10^{28}$ m⁻³s⁻¹ |
| $|H_{DA}|$ | 0.01 eV |
| $\lambda$ | 0.3, 0.4, 0.5, 0.6 eV |

# Supplementary Notes

## *Supplementary note 1: Organic compounds chemical names*

The following organic materials were used as the components in the active layer:

- PM6
  poly[[4,8-bis[5-(2-ethylhexyl)-4-fluoro-2-thienyl]benzo[1,2-*b*:4,5-*b'*]dithiophene-2,6-diyl]-2,5-thiophenediyl-[5,7-bis(2-ethylhexyl)-4,8-dioxo-4*H*,8*H*-benzo[1,2-*c*:4,5-*c'*]dithiophene-1,3-diyl]-2,5-thiophenediyl]

- Y6
  2,2'-[[12,13-bis(2-ethylhexyl)-12,13-dihydro-3,9-diundecylbisthieno[2'',3'':4',5']thieno-[2',3':4,5]pyrrolo[3,2-*e*:2',3'-*g*][2,1,3]benzothiadiazole-2,10-diyl]bis[methylidyne(5,6-difluoro-3-oxo-1*H*-indene-2,1(3*H*)-diylidene)]]bis[propanedinitrile]

- Y5
  2,2'-[[12,13-bis(2-ethylhexyl)-12,13-dihydro-3,9-diundecylbisthieno[2'',3'':4',5']thieno-[2',3':4,5]pyrrolo[3,2-*e*:2',3'-*g*][2,1,3]benzothiadiazole-2,10-diyl]bis[methylidyne(3-oxo-1*H*-indene-2,1(3*H*)-diylidene)]]bis[propanedinitrile]

- o-IDTBR
  (5*Z*,5'*Z*)-5,5'-((7,7'-(4,4,9,9-tetraoctyl-4,9-dihydro-*s*-indaceno[1,2-*b*:5,6-*b'*]dithiophene-2,7-diyl)bis(benzo[*c*][1,2,5]thiadiazole-7,4-diyl))bis(methanylylidene))bis(3-ethyl-2-thioxothiazolidin-4-one)

- PTQ10
  poly[[6,7-difluoro[(2-hexyldecyl)oxy]-5,8-quinoxalinediyl]-2,5-thiophenediyl]]

## *Supplementary Note 2: Constructing bias-dependent trPL quenching*

To construct a bias-dependent TrPL quenching, we use the exciton splitting efficiency $\eta_{LE,split}$ as follows:

$$Max\ of\ \phi_{TrPL}(V) = (1 - \eta_{LE,split}) \cdot Max\ of\ \phi_{TrPL}(0)$$

$$where\ \eta_{LE,split} = 1 - \frac{\int \phi_{TrPL}(V)}{\int \phi_{TrPL}(0)}$$

Then, the maximum of $\phi_{TrPL}(V)$ is normalised and scaled to (1-IGE), so that the normalised emission extrapolated to zero field equals a generation efficiency of EGE=0. This would be the case if all photogenerated NFA singlet excitons would decay radiatively without contributing to free charge generation.

## *Supplementary Note 3: Quantum mechanical calculations*

The natural transition orbitals (NTOs) of the electrons and holes of the excited states along with the excitation energies ($E_{LE}/E_{CT}$), oscillator strength (f) and weight of CT character (CT%) of the Yx+PM6 interfacial dimer system are shown in Figure S21, and the excited state energies for all computed interfacial systems (dimer and trimer) are provided in Table S2 in the supplementary data.

We evaluated the forward and backward reorganization energies ($\lambda_1$, $\lambda_2$) of the interface systems for their NE to CT transition were calculated with the following equation[6]:

$\boldsymbol{\lambda}_1 = E_{eA}(Yx) - E_{eE}(Yx) + E_{nC}(PM6) - E_{nN}(PM6)$,

$\boldsymbol{\lambda}_2 = E_{aE}(Yx) - E_{aA}(Yx) + E_{cN}(PM6) - E_{cC}(PM6)$,

where the $E_{eA}$ and $E_{eE}$ are the Yx excited state energies at the anionic state and excited state optimized geometries respectively, and the $E_{aE}$ and $E_{aA}$ are the Yx anionic state energies at the excited state and anionic state geometries respectively, the $E_{nC}$ and $E_{nN}$ are the PM6 ground state energies at the cationic state and ground state optimized geometries respectively, and the $E_{cN}$ and $E_{cC}$ are the PM6 cationic state energies at the ground state and cationic state geometries respectively. The inner reorganisation energy $\boldsymbol{\lambda}_i$ is then obtained as the average of the forward and backward reorganization energies.

The energies used in the calculation of the forward and backward reorganisation energies are provided in Table S3 in the appendix. The units are in Hartree, which can be converted to eV by multiplying with a factor of 27.21.

_Supplementary Note 4: Comparison of Marcus rate calculation with experimental data_

| System | -ΔIE (eV) | β (-ΔIE + ΔE_b + ΔE_band bending) | -ΔIE + β (eV) |
|---|---|---|---|
| L:Y5 | -0.21 | 0.24 | 0.03 |
| L:Y6 | -0.26 | 0.24 | -0.02 |
| H:Y5 | -0.3 | 0.24 | -0.06 |
| H:Y6 | -0.35 | 0.24 | -0.11 |
| PTQ10:Y5 | -0.18* | 0.24 | 0.06 |
| PTQ10:Y6 | -0.3* | 0.24 | -0.06 |
| PM6:o-IDTBR | - | - | -0.025** |

The above table summarises the scaling of the experimental IE offsets, in order to compare the experimental generation and singlet emission efficiencies with the simulated counterparts from the steady-state rate model based on Marcus theory.

*The ΔIE for PTQ10-based systems was obtained by considering the IE levels of the NFAs in H:Yx blends and subtracting this from a reduced IE level of PM6 in the H:Y5 blend. The IE of PM6 was reduced by 0.12 eV which is the average energetic difference between PTQ10 and PM6, based on CV, PES and PESA measurements in literature.[7]

** For PM6:o-IDTBR, the CV values for this blend from literature suggest an energetic offset even lower than the L:Y5 or PTQ10:Y5 blends.[4] This would, after considering binding energy and band bending effects, would lead the CT state lying nearly 0.3 eV above the LE state, which would not explain the relatively higher generation efficiencies in PM6:o-IDTBR. Hence, we used the LE-CT offset of 0.025 eV (LE above CT), which was previously obtained from the inspection of thermal-activation of the ELQY of this system.[8]

# Appendix Note

By setting the time derivative in both rate equations (eqs. 3a and 3b in the main text) to zero and solving for CT population, the $\eta_{gen,CS}$ is calculated using eq. 5a in the main text, producing the following:

$$\eta_{gen,CS} = \frac{k_{diss,LE}k_{diss,CT}}{(k_{diss,LE} + k_{f,LE})(k_{ref,LE} + k_{f,CT} + k_{diss,CT}) - k_{diss,LE}k_{ref,LE}}$$

### *Test cases:*

1) For the case of fast CT dissociation, $k_{ref,LE} + k_{f,CT} + k_{diss,CT} \cong k_{diss,CT}$. Then,

$$\eta_{gen,CS} \cong \frac{k_{diss,LE}k_{diss,CT}}{k_{diss,LE}k_{diss,CT} + k_{f,LE}k_{diss,CT} - k_{diss,LE}k_{ref,LE}}$$

$$\cong \frac{k_{diss,LE}k_{diss,CT}}{k_{diss,LE}k_{diss,CT} + k_{f,LE}k_{diss,CT}} = \frac{k_{diss,LE}}{k_{diss,LE} + k_{f,LE}}$$

The right-hand side of the above equation denotes the LE dissociation efficiency competing with its decay. So $\eta_{gen,CS}$ is determined by $k_{diss,LE}(\Delta E, \lambda)$ and progressively depends on $\lambda$, seen in the progression of the sub-figures in SI Figure S22 towards fast CT dissociation.

In other words, $\eta_{gen,CS}$ in the transition region has the same shape as $k_{diss,LE}$. We see this in the Figure 7d of the main text.

2) Slow CT dissociation (or fast LE reformation), so $k_{ref,LE} + k_{f,CT} + k_{diss,CT} \cong k_{ref,LE}$. Then,

$$\eta_{gen,CS} \cong \frac{k_{diss,LE}k_{diss,CT}}{k_{diss,LE}k_{ref,LE} + k_{f,LE}k_{ref,LE} - k_{diss,LE}k_{ref,LE}}$$

$$= \frac{k_{diss,LE}}{k_{f,LE}}\frac{k_{diss,CT}}{k_{ref,LE}} = \frac{k_{diss,CT}}{k_{f,LE}}\frac{k_{diss,LE}}{k_{ref},LE}$$

Here, the right-hand side of the equation contains $\frac{k_{diss}}{k_{ref}}$, whose ratio of coefficients is independent of $\lambda$ (see SI Figure S23) but depends on $\Delta E$. Hence, in the scenario when a slow CT dissociation establishes a kinetic equilibrium between the CT and LE population, the free charge generation efficiency shows a reduced dependence on the reorganisation energy, which we see this in Figure S22a for slow CT dissociation.

3) Fast LE dissociation, so $k_{diss,LE} + k_{f,LE} \cong k_{diss,LE}$. Then,

$$\eta_{gen,CS} \cong \frac{k_{diss,LE}k_{diss,CT}}{k_{diss,LE}(k_{ref,LE} + k_{f,CT} + k_{diss,CT}) - k_{diss,LE}k_{ref,LE}} \cong \frac{k_{diss,CT}}{k_{f,CT} + k_{diss,CT}}$$

Here, the right-hand side of the equation denotes the CT dissociation efficiency competing with CT decay which has a smaller dependence on the reorganization energy than the previous test cases. We see this in SI Figure S24.